\documentclass[lettersize,journal]{IEEEtran}
\usepackage{amsmath,amsfonts}
\usepackage{algorithmic}
\usepackage{algorithm}
\usepackage{array}
\usepackage[caption=false,font=normalsize,labelfont=sf,textfont=sf]{subfig}
\usepackage{textcomp}
\usepackage{stfloats}
\usepackage{url}
\usepackage{verbatim}
\usepackage{graphicx}
\usepackage{cite}
\usepackage{multirow}
\usepackage{makecell}
\usepackage{threeparttable}
\usepackage{url}
\usepackage[colorlinks, linkcolor=red, anchorcolor=red, citecolor=green]{hyperref}
\usepackage{color}

\begin{document}
\setcounter{table}{0}
\title{Self-Supervised Learning of Spatial Acoustic Representation with Cross-Channel Signal Reconstruction and Multi-Channel Conformer}

\author{Bing~Yang and Xiaofei~Li
\thanks{Bing Yang and Xiaofei Li are with the School of Engineering, Westlake University, Hangzhou 310030, China, and also with the Institute of Advanced Technology, Westlake Institute for Advanced Study, Hangzhou 310024, China (e-mail: yangbing@westlake.edu.cn; lixiaofei@westlake.edu.cn). (Corresponding author: Xiaofei Li.)}
}



\maketitle

\begin{abstract}
Supervised learning methods have shown effectiveness in estimating spatial acoustic parameters such as time difference of arrival, direct-to-reverberant ratio and reverberation time. However, they still suffer from the simulation-to-reality generalization problem due to the mismatch between simulated and real-world acoustic characteristics and the deficiency of annotated real-world data. To this end, this work proposes a self-supervised method that takes full advantage of unlabeled data for spatial acoustic parameter estimation. First, a new pretext task, i.e. cross-channel signal reconstruction (CCSR), is designed to learn a universal spatial acoustic representation from unlabeled multi-channel microphone signals. We mask partial signals of one channel and ask the model to reconstruct them, which makes it possible to learn spatial acoustic information from unmasked signals and extract source information from the other microphone channel. An encoder-decoder structure is used to disentangle the two kinds of information. By fine-tuning the pre-trained spatial encoder with a small annotated dataset, this encoder can be used to estimate spatial acoustic parameters. Second, a novel multi-channel audio Conformer (MC-Conformer) is adopted as the encoder model architecture, which is suitable for both the pretext and downstream tasks. It is carefully designed to be able to capture the local and global characteristics of spatial acoustics exhibited in the time-frequency domain. Experimental results of five acoustic parameter estimation tasks on both simulated and real-world data show the effectiveness of the proposed method. To the best of our knowledge, this is the first self-supervised learning method in the field of spatial acoustic representation learning and multi-channel audio signal processing.
\footnote{Codes are available at \url{https://github.com/Audio-WestlakeU/SAR-SSL}.}

\end{abstract}

\begin{IEEEkeywords}
Multi-channel audio signal processing, self-supervised learning, spatial acoustic parameter estimation. 
\end{IEEEkeywords}

\section{Introduction}
\IEEEPARstart{S}{patial} acoustic representation learning aims to extract a low-dimension representation of the spatial propagation characteristics of sound from microphone recordings. It can be used in a variety of audio tasks, such as estimating the acoustic parameters \cite{ACE16} or the geometrical information \cite{GEO_REF_fromRIR21} related to source, microphone and environment.
Related audio tasks have been widely applied in augmented reality \cite{VR15} and hearing aids \cite{Hearingaid01} where generating perceptually acceptable sound for the target environment is required to guarantee a good immersive experience, and also in intelligent robots \cite{Robot21} where perceiving surrounding acoustic properties serves as a priori for robot interaction with humans and environments.

Room impulse responses (RIRs) characterize the sound propagation from sound source to microphone constrained in a room environment. The physically relevant parameters that determine RIR include the positions of sound source and microphone array, the room geometry, and the absorption coefficients of walls. RIR is composed of direct-path propagation, early reflections, and late reverberation.
The relative position between the source and the microphone array determines the direct-path propagation. All the three physical parameters affect the arrival times and the strengths of reflective pulses including early reflections and late reverberation. Some acoustic parameters of the spatial environment can be directly estimated from RIR without supervision, like position-dependent parameters including time difference of arrival (TDOA), direct-to-reverberant ratio (DRR) and clarity index $C_{50}$, or position-independent parameters including reverberation time $T_{60}$. The physical parameters including absorption coefficient, surface area, and room volume are difficult to estimate by conventional signal processing techniques since the RIR can not be formulated as an analytical function of them. Some researchers try to estimate these parameters from RIRs with deep neural networks (DNNs) \cite{ABS_CNNMLP_fromRIR_JASA21, GEO_REF_fromRIR21}. However, the RIR is normally unavailable without intrusive measurement in practical applications, in which case some works predict RIR \cite{RIR_ICASSP22, RIR_ICASSP23} or spatial acoustic embedding \cite{RIREmbed_ICASSP20, T60_C50_VOL_ICASSP23} blindly from microphone recordings. As environmental acoustic parameters and geometrical information are heavily relevant to RIRs, a good blind pre-predictor of RIR or spatial acoustic representation will definitely benefit further spatial acoustic parameter estimation and geometrical structure analysis.

With the development of deep learning techniques, lots of works directly estimate spatial acoustic parameters from microphone signals in a supervised manner. These supervised works have achieved superior performance than conventional methods, owing to the strong modeling ability of deep neural networks. Since these works are data-driven, the diversity and quantity of training data are crucial to their performance. To this end, DNN models are usually trained with abundant diverse simulated data, and then transferred to real-world data.
Some researchers show that the trained model does not perform well when directly transferred to real-world data \cite{T60_SUR_VOL_IWAENC22} due to the mismatch between simulated and real-world RIRs. In \cite{T60_SUR_VOL_IWAENC22}, the mismatch is analyzed mainly in terms of the directivity of source and microphone, and the wall absorption coefficient. Moreover, there are many other aspects of mismatch, for example:
1) The acoustic response of real-world moving source \cite{Diffuse_ISM10} and the spatial correlation of real-world multi-channel noise are difficult to simulate.
2) RIR simulators \cite{gpuRIR20} usually generate empty box-shaped rooms, while obstacles and non-regular-shaped rooms exist in real-world applications.
As an alternative solution, real-world data can be also used for training. However, annotating the acoustic environment and the sound propagation paths would be very difficult and expensive. Existing annotated real-world datasets lack diversity and quantity, which limits the development of supervised learning methods. Therefore, it is necessary to study how to dig spatial acoustic information from unlabeled real-world data.

In this work, we investigate how to learn a universal spatial acoustic representation from unlabeled dual-channel microphone signals based on self-supervised learning.
Microphone signals can be formulated as a convolution between dry source signals with multi-channel RIRs, with the addition of noise signals. Spatial acoustic representation learning focuses on extracting RIR-related but source-independent embeddings. As far as we know, this is the first work studying on self-supervised learning of spatial acoustic representation.
The proposed method has the following contributions.

\subsubsection{Self-supervised learning of spatial acoustic representation (SSL-SAR)} The proposed method follows the basic pipeline of self-supervised learning, namely first pre-training using abundant unlabeled data and then fine-tuning using a small labeled dataset for downstream tasks. A new pretext task, i.e. cross-channel signal reconstruction (CCSR), is designed for self-supervised learning of a universal spatial acoustic representation. This work is implemented in the short-time Fourier transform (STFT) domain. Given the dual-channel microphone signals as input, we randomly mask a portion of STFT frames of one microphone channel and ask the neural network to reconstruct them.
The reconstruction of the masked STFT frames requires both the spatial acoustic information related to RIRs and the spectral pattern information indicating source signal content.
Accordingly, the network is forced to learn the (inter-channel) spatial acoustic information from the frames that are not masked for both microphone channels, and meanwhile extract the spectral information from corresponding frames of the unmasked channel.
In order to disentangle the two kinds of information, the input STFT coefficients are separately masked and fed to two different encoders. The spatial and spectral representations are concatenated along the embedding dimension and then passed to a decoder to reconstruct the masked STFT frames. The pre-trained spatial encoder can provide useful information to various spatial acoustics-related downstream tasks.
Note that this work only considers static acoustic scenarios where RIRs are time-invariant.

\subsubsection{Multi-channel audio Conformer (MC-Conformer)}
Since the network would be pre-trained according to the pretext task, and then adopted by various downstream tasks, we need a powerful network that is suitable for both pretext and downstream tasks.
To this end, a novel MC-Conformer is adopted as the encoder model. To fully learn the local and global properties of spatial acoustics exhibited in the time-frequency (TF) domain, it is designed following a local-to-global processing pipeline.
The local processing model applies 2D convolutional layers to the raw dual-channel STFT coefficients to learn the relationship between microphone signals and RIRs. It captures the short-term and sub-band spatial acoustic information.
The global processing model uses Conformer \cite{Conformer20} blocks to mainly learn the full-band and long-term relationship of spatial acoustics. The feed-forward modules of Conformer can model the full-band correlations of RIRs, namely the wrapped-linear correlation between frequency-wise IPDs and time-domain TDOA for the direct path and early reflections. Considering RIRs are time-invariant for the entire signal, the multi-head self-attention module is used to model such long-term temporal dependence.
Though the combination of 2D CNN and Conformer has been used in other tasks such as sound event localization and detection \cite{CNN-Conformer21}, in this work, we investigate its use for spatial acoustic representation learning and spatial acoustic parameter estimation.

The rest of this paper is organized as follows. Section \ref{sec:related_work} overviews the related works in the literature. Section \ref{sec:problem_formulation} formulates the spatial acoustic representation learning problem. Section \ref{sec:method} details the proposed self-supervised spatial acoustic representation learning method. Experiments and discussions with simulated and real-world data are presented in Section \ref{sec:exp}, and conclusions are drawn in Section \ref{sec:conclusion}.

\vspace{-0.1cm}
\section{Related works}
\vspace{-0.1cm}
\label{sec:related_work}
\begin{table*}[]
    \centering
    \caption{Summary of deep-learning-based spatial acoustic parameter estimation methods}
    \label{tab:methodoverview}
    \renewcommand\arraystretch{1.2}
    \tabcolsep0.04in
    \begin{threeparttable}
    \begin{tabular}{lccccccccc}
        \hline
        \hline
        \multirow{2}{*}{Method} &\multirow{2}{*}{Estimated Parameter} &\multirow{2}{*}{\# Mic.} &\multirow{2}{*}{Input} &\multirow{2}{*}{Model} &\multirow{2}{*}{Supervision} \\
        & \\
        \hline
        \cite{TDOA_LSTM_ICASSP19} &TDOA &Two &Log-mel magnitude spectrogram, GCC-PHAT &LSTM &Supervised \\
        \cite{YBTASLP21} &DOA &Two &Log magnitude and phase spectrograms &CRNN &Supervised \\
        \cite{WYBIS23} &DOA &Two &Real and imaginary spectrograms &LSTM &Supervised \\
        \cite{DRR_LSTM_IS23} &DRR &One &Log magnitude spectrogram &LSTM &Supervised\\
        \cite{T60_CNN_IWAENC18} &$T_{60}$ &One &Log-gammatone energy spectrogram &CNN &Supervised\\
        \cite{T60_CRNN_IS20} &$T_{60}$ &One &Log-gammatone energy spectrogram &CRNN &Supervised\\
        \cite{ABS_CNNMLP_fromRIR_JASA21} &Absorption coefficient &One &RIR &CNN, MLP &Supervised\\
        \cite{VOL_CNN_ICASSP19} &Room volume &One &Log-gammatone energy spectrogram, energy-based features &CNN &Supervised\\
        \cite{T60_ELR_TASLP19} &$T_{60}$, $C_{50}$  &One &Log-gammatone temporal-modulated spectrogram &MLP  &Supervised\\
        \cite{DRR_T60_SNR_ICASSP20} &DRR, $T_{60}$, SNR &One &Magnitude spectrogram &CNN &Supervised\\
        \cite{SNR_STI_T60_C50_C80_DRR_WASPAA21} &$T_{60}$, DRR, $C_{50}$, $C_{80}$, SNR, STI &One &MFCC &CRNN &Supervised\vspace{0.1cm}\\
        \cite{ABS_T60_SUR_VOL_WASPAA21} &\makecell{$T_{60}$, absorption coefficient, \\surface area, room volume} &Two &Magnitude spectrogram, ILD, IPD &CNN &Supervised\vspace{0.1cm}\\
        \cite{T60_SUR_VOL_IWAENC22} &$T_{60}$, surface area, room volume &Two &Magnitude spectrogram, ILD, IPD &CNN &Supervised\\
        \cite{VOL_T60_ICASSP23} &$T_{60}$, room volume &One &\makecell{Log-gammatone energy and phase spectrograms, energy-based features} &CNN &Supervised\\
        \cite{T60_C50_JASA23} &$T_{60}$, $C_{50}$ &One &Log-gammatone magnitude spectrogram &CRNN &Supervised\\
        \cite{T60_C50_VOL_ICASSP23} &$T_{60}$, $C_{50}$, room volume &One &Log magnitude spectrograms &CNN &Supervised\vspace{0.1cm}\\
        \hline
        \multirow{2}{*}{Proposed} &TDOA, DRR, $T_{60}$, $C_{50}$, &\multirow{2}{*}{Two} &\multirow{2}{*}{Real and imaginary spectrograms} &\multirow{1}{*}{CNN, } &\multirow{2}{*}{Self-supervised} \\
        &absorption coefficient & & &Conformer\\
        \hline
        \hline
    \end{tabular}
    \end{threeparttable}
\end{table*}

\subsection{Deep-Learning-Based Spatial Acoustic Parameter Estimation}
Spatial acoustic parameter estimation can provide important acoustic information of the environment. Lots of deep learning based methods are developed for related tasks \cite{TDOA_LSTM_ICASSP19,YBTASLP21,WYBIS23,DRR_LSTM_IS23,T60_CNN_IWAENC18,T60_CRNN_IS20,ABS_CNNMLP_fromRIR_JASA21,VOL_CNN_ICASSP19,T60_ELR_TASLP19,DRR_T60_SNR_ICASSP20,SNR_STI_T60_C50_C80_DRR_WASPAA21,ABS_T60_SUR_VOL_WASPAA21,T60_SUR_VOL_IWAENC22,VOL_T60_ICASSP23,T60_C50_JASA23,T60_C50_VOL_ICASSP23}, which are summarised in Table I.
The estimation of spatial location-related parameters like TDOA and direction of arrival (DOA) requires multi-channel microphone signals as input \cite{TDOA_LSTM_ICASSP19,YBTASLP21,WYBIS23}. The other spatial acoustic parameters can be predicted with both single-channel \cite{DRR_LSTM_IS23,T60_CNN_IWAENC18,T60_CRNN_IS20,ABS_CNNMLP_fromRIR_JASA21,VOL_CNN_ICASSP19,T60_ELR_TASLP19,DRR_T60_SNR_ICASSP20,SNR_STI_T60_C50_C80_DRR_WASPAA21,VOL_T60_ICASSP23,T60_C50_JASA23,T60_C50_VOL_ICASSP23} and multi-channel microphone signals \cite{ABS_T60_SUR_VOL_WASPAA21,T60_SUR_VOL_IWAENC22}, such as DRR, $C_{50}$, $T_{60}$, absorption coefficient, total surface area, room volume, signal-to-noise ratio (SNR) and speech transmission Index (STI).
The network input can be spectrograms involving magnitude/energy \cite{TDOA_LSTM_ICASSP19,YBTASLP21,DRR_LSTM_IS23,T60_CNN_IWAENC18,T60_CRNN_IS20,VOL_CNN_ICASSP19,DRR_T60_SNR_ICASSP20,ABS_T60_SUR_VOL_WASPAA21,T60_SUR_VOL_IWAENC22,VOL_T60_ICASSP23,T60_C50_JASA23,T60_C50_VOL_ICASSP23}, phase \cite{T60_C50_JASA23}, real and imaginary parts \cite{WYBIS23} of the complex-valued STFT coefficients, or relatively high-level features like inter-channel level difference (ILD) \cite{ABS_T60_SUR_VOL_WASPAA21,T60_SUR_VOL_IWAENC22}, inter-channel phase difference (IPD) \cite{ABS_T60_SUR_VOL_WASPAA21,T60_SUR_VOL_IWAENC22}, generalized cross-correlation function with phase transform (GCC-PHAT) \cite{TDOA_LSTM_ICASSP19}, mel-scale frequency cepstral coefficient (MFCC) \cite{SNR_STI_T60_C50_C80_DRR_WASPAA21} and energy-based features \cite{VOL_CNN_ICASSP19,T60_C50_JASA23}. The commonly used model architectures include long short-term memory (LSTM) model \cite{TDOA_LSTM_ICASSP19,WYBIS23,DRR_LSTM_IS23}, convolutional neural network (CNN) \cite{T60_CNN_IWAENC18,ABS_CNNMLP_fromRIR_JASA21,VOL_CNN_ICASSP19,T60_ELR_TASLP19,DRR_T60_SNR_ICASSP20,ABS_T60_SUR_VOL_WASPAA21,T60_SUR_VOL_IWAENC22,VOL_T60_ICASSP23,T60_C50_VOL_ICASSP23}, convolutional recurrent neural network (CRNN) \cite{YBTASLP21,T60_CRNN_IS20,SNR_STI_T60_C50_C80_DRR_WASPAA21,T60_C50_JASA23}, multilayer perceptron (MLP) \cite{ABS_CNNMLP_fromRIR_JASA21,T60_ELR_TASLP19}, etc.

Most existing works train neworks for an acoustic parameter solely or multiple acoustic parameters jointly with labeled data, which implicitly learn task-oriented spatial acoustic information in a fully supervised manner. The work \cite{T60_C50_VOL_ICASSP23} extracts a universal representation of the acoustic environment with a contrastive learning method. However, it requires the RIR annotation to obtain the positive and negative sample pairs, which is still a supervised learning method. In contrast to these works, we aim to design a self-supervised method to learn a universal spatial acoustic representation that can be applied to various spatial acoustic-related downstream tasks. Self-supervised learning does not require any data annotation, which allows intensively digging spatial acoustic information from large-scale unlabeled audio data, especially from the real-world recordings.

\subsection{Audio Self-Supervised Representation Learning}
Self-supervised representation learning \cite{SSL22} has been successfully applied to audio/speech processing in recent years \cite{SSL_speech22}, which has shown effectiveness in a wide range of downstream applications like automatic speech recognition and sound event classification. According to how to build pretext task, it can be grouped into two categories, namely contrastive approaches and generative approaches.
\textit{Contrastive approaches} aim to learn a latent representation space that pulls close positive samples, and sometimes meanwhile pulls away negative samples from positive samples. Typical methods include contrastive predictive coding  \cite{CPC18}, wav2vec \cite{wav2vec19}, COLA \cite{COLA21}, BYOL-Audio \cite{BYOLA23}, etc.
\textit{Generative approaches} learn representation by generating or reconstructing the input audio data with some limited views. Autoregressive predictive coding \cite{APC19,APC20} predicts future inputs from past inputs with an unsupervised autoregressive neural model to learn generic speech representation. Inspired by the masked language model task from BERT \cite{BERT19}, some researchers propose to learn general-purpose audio representation by reconstructing masked patches from unmasked TF regions using Transformer \cite{SSAST22,MAEAST22}.

These methods learn the representation of sound source from single-channel signal, and remove the influence of channel effect such as the response introduced by propagation paths. This kind of representation can be applied to a number of signal-content-related downstream tasks. In contrast, this work aims to learn the representation of spatial acoustic information and remove the information of the sound source, which will be used in spatial-acoustic-related downstream applications.
Though there are some unsupervised/self-supervised methods proposed for multi-channel signal processing \cite{SepLocSCM23, SepUnsup23}, their self-supervised pretext tasks are different from our method. We aim to learn a general spatial acoustic representation, while they are designed for specific tasks and the potential to learn spatial acoustic information is unknown.
Different from the existing masking-reconstruction pretext task \cite{SSAST22,MAEAST22} which encourages learning inner-channel information, the proposed pretext task, i.e. cross-channel signal reconstruction, intends to learn both inner-channel and inter-channel information.

\section{Problem Formulation}
\label{sec:problem_formulation}
We consider the case that one static sound source is observed by two static microphones in an enclosed room environment with additive ambient noise. The signal captured by the $m$-th microphone is denoted as
\begin{equation}
    x_m(t) = h_{m}(t) * s(t) + v_m(t),
  \label{eq_model_T}
\end{equation}
where $m \in \{1, M\}$ denotes the microphone index, $t \in [1, T]$ is the time sample index, $s(t)$ is the source signal, and $v_m(t)$ is the received noise signal at the $m$-th microphone. Here, $M$ is always set to 2, and $*$ denotes convolution.
The RIR $h_{m}(t)$ from the sound source to the $m$-th microphone consists of two successive parts, which is formulated as
\begin{equation}
    h_m(t) = h_{m}^{\rm{d}}(t) + h_{m}^{\rm{r}}(t),
  \label{eq_air}
\end{equation}
where $h_{m}^{\rm{d}}(t)$ and $h_{m}^{\rm{r}}(t)$ are the impulse responses of the direct path and reflected paths, respectively.
Theoretically, the sound arrives at the microphones first along the direct path. The impulse response of the direct path is formulated as
\begin{equation}
    h_{m}^{\rm{d}}(t)=\alpha_m \delta(t-\tau_m),
  \label{eq_dpair}
\end{equation}
where $\alpha_{m}$ and $\tau_{m}$ are the propagation attenuation factor and the time of arrival from the source to the $m$-th microphone, respectively.
The dual-channel direct-path pulses can reflect the DOA of the source relative to the microphone array.
The following pulses relate to the reflections on room boundaries or built-in objects, which can be divided into early reflections and late reverberation on the basis of their arrival times. The reflected paths indicate the acoustic settings of room, e.g., the decaying rate of RIR reflects the $T_{60}$ of room.

\begin{figure}[t]
  \centering
  \includegraphics[width=0.98\linewidth]{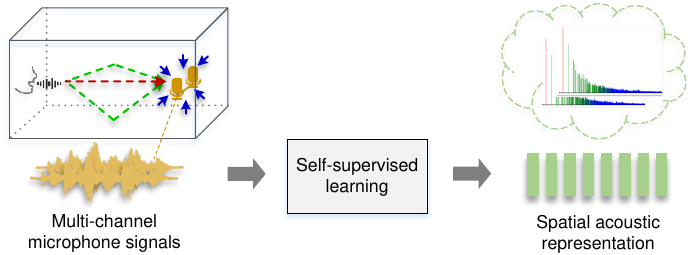}
  \vspace{-0.1cm}
  \caption{Illustration of self-supervised learning of spatial acoustic representation using multi-channel microphone recordings. The direct path, early reflections and late reverberation are illustrated in red, green and blue colors, respectively. }
  \label{fig:meth_task}
\end{figure}

Considering the sound propagation and the reflection when encountering obstacles are frequency-dependent, we 
convert the time-domain signal model in Eq.~\eqref{eq_model_T} into the STFT domain as \cite{talmon2009relative}
\vspace{-0.1cm}
\begin{equation}
    X_m(n,f) \! \approx \! \sum_{n'}{H_m(n', f) S(n-n',f)}  + V_m(n,f),
    \label{eq_model_TF_CTF}
\vspace{-0.1cm}
\end{equation}
where $n \in [1, N]$ and $f \in [1, F]$ represent the time frame index and frequency index, respectively, and $N$ and $F$ are the numbers of frames and frequencies, respectively. Here, $X_m(n,f)$, $S(n,f)$ and $V_m(n,f)$ represent the microphone, source and noise signals in the STFT domain, respectively. In the STFT-domain signal model, RIR can be well approximated by its sub-band representation, namely the convolutive transfer function (CTF) $H_m(n, f)$.

The spatial acoustic information is encoded in the dual-channel RIRs/CTFs, being independent of the source signal and ambient noise.
As illustrated in Fig.~\ref{fig:meth_task}, this work aims to design a self-supervised method to learn a universal spatial acoustic representation related to the RIRs/CTFs, from unlabeled dual-channel microphone signals. The representation can be used to estimate the spatial acoustic parameters including TDOA, DRR, $T_{60}$, $C_{50}$, absorption coefficients, etc.
Dual-channel microphone signals are given as the input of the self-supervised model since the estimation of some position-dependent parameters like TDOA and DOA requires signals of at least two microphones, and dual-channel microphone signals are expected to provide more reliable spatial acoustic information than single-channel signal \cite{ABS_T60_SUR_VOL_WASPAA21}.
Moreover, as will be shown later, using two microphones allows us to design a self-supervised pretext task that is able to better disentangle the spatial acoustic information and source information.
Considering the simplicity of the signal model in Eq.~\eqref{eq_model_TF_CTF}, this work intends to design a self-supervised learning model in the STFT domain, and learn the CTF information from $X_m(n,f)$.
Though some parameters like DRR, $T_{60}$, and absorption coefficients are frequency-dependent, this work focuses on estimating the full-band versions of them \cite{ACE16} which are obtained by averaging across the sub-band values.

\section{Self-Supervised Learning of Spatial Acoustic Representation}
\label{sec:method}

\begin{figure}[t]
  \centering
  \includegraphics[width=0.99\linewidth]{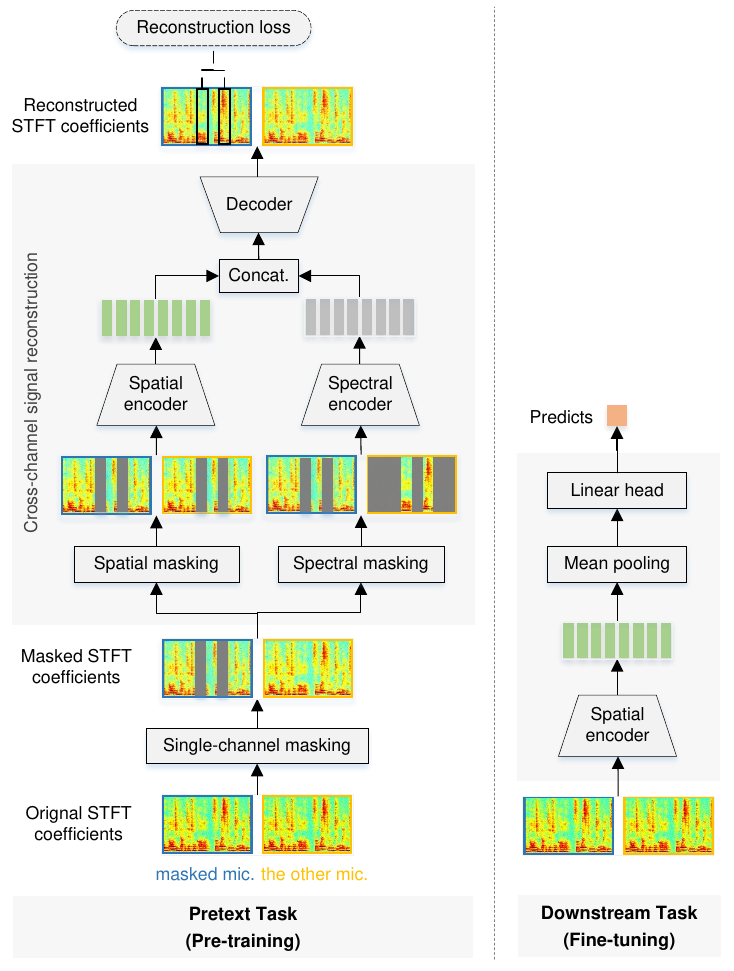}
  \caption{Block diagram of the proposed self-supervised spatial acoustic representation learning model. The complex-valued STFT coefficients are illustrated by their real-part spectrograms.}
  \label{fig:meth_flowchart}
\end{figure}

The proposed spatial acoustic representation learning method follows the basic pipeline of most self-supervised learning methods, namely first pre-training the representation model according to the pretext task using a large amount of unlabeled data, and then fine-tuning the pre-trained model for a specific downstream task using a small amount of labeled data.
The key points of this work lie in how to build the pretext task to learn spatial acoustic information (see details in Section \ref{sec:reconstruct}), and how to design a unified network architecture suited for both pretext task and downstream tasks (see details in Section \ref{sec:mc_conformer}).
The block diagram of the proposed method is shown in Fig.~\ref{fig:meth_flowchart}.

\subsection{Pretext Task: Cross-Channel Signal Reconstruction}
\label{sec:reconstruct}
A cross-channel signal reconstruction (CCSR) pretext task is built to learn the spatial acoustic information. As illustrated in Fig.~\ref{fig:meth_flowchart}, the basic idea is to mask a portion of STFT frames of one microphone channel to destroy corresponding spectral and spatial information, and then ask the network to reconstruct them.
The expected function of the reconstruction network lies in three aspects:

i) Learning the \textbf{spectral patterns} of masked frames from the unmasked channel.
Source signals have unique spectral patterns indicating the signal content. Since signals received at the two microphones have the same spectral information, we only mask one channel to preserve the corresponding signal content, and the network can learn the spectral information of sound source from the unmasked channel. The learned information from the unmasked channel may also include some RIR/CTF information of the unmasked channel.

ii) Learning the \textbf{spatial acoustics} from the dual-channel unmasked frames. To reconstruct the masked STFT frames, the network needs to learn the (inter-channel) acoustic information from the dual-channel unmasked frames, and apply it to the information learned from the unmasked channel. The (inter-channel) spatial acoustic information relates to the RIR/CTF of the masked channel and more possibly to the relative RIR/CTF of the masked channel to the unmasked channel. In the representation of relative CTF/RIR, the relative information can be directly used for inter-channel downstream tasks, such as TDOA estimation. Moreover, it is expected that the temporal structure of CTF/RIR (or a variant of the temporal structure of CTF/RIR) is also preserved, from which the information used for temporal-structure-related downstream tasks, such as $T_{60}$ estimation, can be extracted by DNN mapping. These assumptions will be validated through experiments, in which the learned spatial acoustic representation is shown to be effective for a variety of downstream tasks.

iii) Reconstructing the masked frames using the learned spectral and spatial information.

The proposed cross-channel signal reconstruction pretext task can disentangle source spectral information and spatial information, which facilitates the application of spatial acoustic representation in downstream tasks.

\subsubsection{Reconstruction framework}
A portion of the STFT frames of one microphone channel is randomly masked. The signal masked by the single-channel masking operation is formulated as
\begin{equation}
    \tilde{X}_{m_{\rm{mask}}}(n,f) = X_{m_{\rm{mask}}}(n,f)W(n,f),
    \label{eq_mask_encoder}
\end{equation}
where $m_{\rm{mask}}$ denotes the index of the masked channel, which is randomly selected from the two microphones for each training sample. $W(n,f)$ is the binary-valued TF mask, and 0 for masking while 1 for not masking.
The pretext task is to reconstruct the STFT coefficients of masked frames given the STFT coefficients of both the masked and the unmasked channels.
The mean squared error (MSE) between the original and the reconstructed STFT coefficients of the masked frames is adopted as the reconstruction loss, which is formulated as
\begin{equation}
\small
    L = \frac{\sum_{n=1,f=1}^{N,F}\left|X_{m_{\rm{mask}}}(n,f)-\hat{X}_{m_{\rm{mask}}}(n,f)\right|^2(1-{W}(n,f)) }{\sum_{n=1,f=1}^{N,F}(1-{W}(n,f))},
    \label{eq_loss}
\end{equation}
where $\hat{X}_{m_{\rm{mask}}}(n,f)$ is the reconstructed STFT coefficients, and $|\cdot|$ denotes the magnitude of the complex number. Here, $1-W(n,f)$ indicates that the reconstruction loss is only computed on the masked frames.

\subsubsection{Encoder-decoder structure}
\begin{figure}[t]
  \centering
  \includegraphics[width=0.98\linewidth]{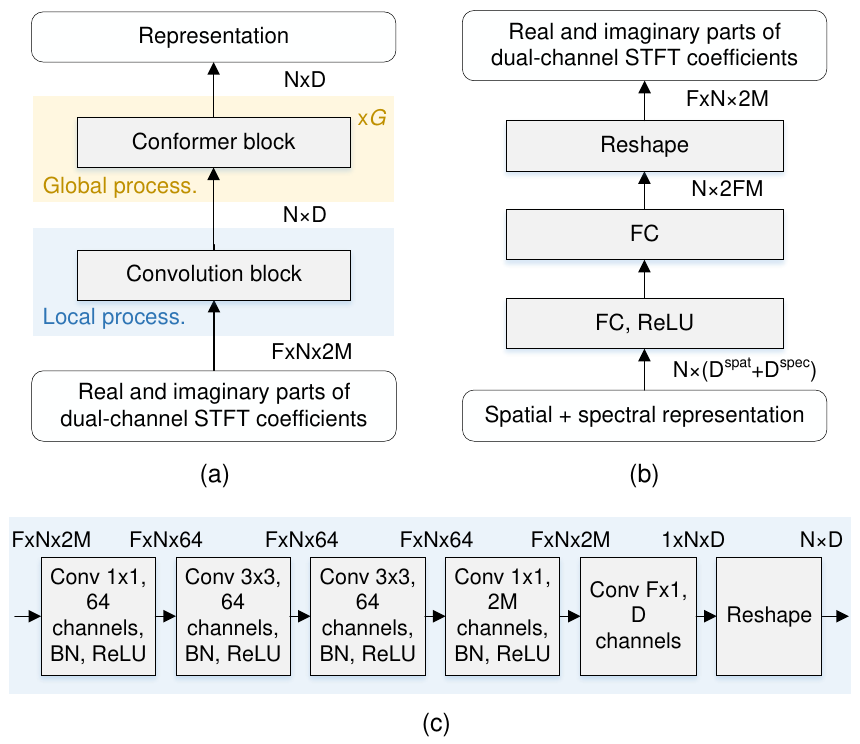}
  \caption{Model architecture of (a) spatial/spectral encoder (namely multi-channel audio Conformer), (b) decoder and (c) convolution block in the encoder. $D$ is the hidden dimension, and $D=D^{\rm{spat}}$ for spatial encoder and $D=D^{\rm{spec}}$ for spectral encoder. }
  \label{fig:meth_network}
\end{figure}

The model of the pretext task adopts an encoder-decoder structure, as shown in Fig.~\ref{fig:meth_flowchart}. Considering the characteristics and heterogeneity of spatial acoustics and signal content, the input STFT coefficients are first masked in two different ways (see details in Section \ref{sec:mask})), then fed into spatial and spectral encoders to separately learn the two kinds of information. Both encoders adopt the MC-Conformer (see details in Section \ref{sec:mc_conformer}) but with different configurations (see details in Section \ref{sec:model_configuration}).
The complex-valued STFT coefficients have a dimension of $F \times N \times M$. The concatenation of their real and imaginary parts along the channel dimension is taken as the input of encoders, which has a dimension of $F \times N \times 2M$. The two encoders convert their own masked inputs to a spatial embedding sequence of $N \times D^{\rm{spat}}$ and a spectral embedding sequence of $N \times D^{\rm{spec}}$, respectively. $D^{\rm{spat}}$ and $D^{\rm{spec}}$ are the hidden dimensions.
Note that this work intends to learn the spatial acoustic information, and the learning of spectral information is just designed for disentangling the spectral information from the learning of spatial acoustic representation and making the spatial acoustic representation learning more concentrated.
The embeddings learned by the two encoders are concatenated along the hidden dimension and then fed into the decoder to predict the real and imaginary parts of the dual-channel STFT coefficients. Though the reconstruction loss is only calculated on the masked frames of one channel, we still output the dual-channel STFT coefficients to preserve the frame and channel information of the original input.
As shown in Fig.~\ref{fig:meth_network} (b), two fully-connected (FC) layers are adopted as the decoder, and each frame is separately processed by the same decoder. In order to encourage the spatial and spectral information used for signal reconstruction to be fully learned by the encoders instead of the decoder, the decoder is set to not have any information interaction between time frames.

\subsubsection{Masking scheme}
\label{sec:mask}
The masked signal $\tilde{X}_{m_{\rm{mask}}}(n,f)$ is readjusted for the two encoders to disentangle spatial acoustics information and signal content information.
In order to encourage the spatial encoder to focus on learning the spatial acoustic information, we remove the spectral information of masked frames from the input of the spatial encoder. Specifically, we apply the single-channel mask to both microphone channels, which is formulated as
\begin{equation}
    X_m^{\rm{spat}}(n,f) = X_m(n,f) W(n,f).
    \label{eq_mask_spat}
\end{equation}
It guarantees that the spatial encoder will not see any channel of the masked frames, and hopefully will focus on learning the spatial information from unmasked frames.

The spectral encoder is used to learn signal content information indicated by the inner-channel information.
To make the spectral encoder learn the signal content information, we only input the single-channel signal to the spectral encoder. Specifically, the inverse single-channel mask is additionally applied to the signals of the other microphone, which is formulated as
\begin{equation}
    X_m^{\rm{spec}}(n,f) =
    \left\{
    \begin{aligned}
        &X_m(n,f) W(n,f), \ \text{if} \ m= m_{\rm{mask}}\\
        &X_m(n,f)( 1- W(n,f)), \  \rm{otherwise}
    \end{aligned}
    \right..\\
    \label{eq_mask_spec}
\end{equation}
The spectral encoder cannot see the same frames of both microphones simultaneously, and thus will not learn the inter-channel spatial acoustic information. To model the spectral information of masked frames, in addition to the corresponding frames of unmasked channel, we also input the unmasked frames of the masked channel to provide more context information for the masked frames.
The spectral encoder may also learn some single-channel RIR/CTF information.

\subsection{Downstream Tasks: Spatial Acoustic Parameter Estimation}
Fig.~\ref{fig:meth_flowchart} also shows the diagram of how to use the pre-trained model in downstream tasks. The dual-channel STFT coefficients $X_m(n,f)$ are directly passed to the spatial encoder without any masking operation.
The spatial encoder processes the input and produces a spatial acoustic representation. This representation is then pooled across all time frames and passed to a linear head to estimate the spatial acoustic parameter.
The weights of the spatial encoder are initialized from the self-supervised pre-trained model and then fine-tuned using the labeled data of downstream tasks.

We consider estimating the following spatial acoustic parameters as downstream tasks.
\begin{itemize}
    \item TDOA is an important feature for sound source localization. It is defined as the relative time delay, namely $\Delta t$ in seconds, when the sound emitted by the source arrives at the two microphones.
    This work estimates TDOA in samples, namely $\Delta t f_s$, where $f_s$ is the sampling rate of signals.
    \item DRR is defined as the energy ratio of direct-path part to the rest of RIRs \cite{ACE16}, i.e.,
    \begin{equation}
       {\rm{DRR}_m} = 10\log_{10} \frac{\sum_{t=t_{\rm{d}}-\Delta t_{\rm{d}}}^{t_{\rm{d}}+\Delta t_{\rm{d}}}h^2_m(t)}{{\sum_{t=0}^{t_{\rm{d}}-\Delta t_{\rm{d}}}h^2_m(t)+{\sum_{t=t_{\rm{d}}+\Delta t_{\rm{d}}}^{\infty}h^2_m(t)}}},
    \label{eq_drr}
    \end{equation}
    where the direct-path signal arrives at the $t_{\rm{d}}$-th sample, and $\Delta t_{\rm{d}}$ is the additional sample spread for the direct-path signal, which typically corresponds to 2.5 ms \cite{ACE16}.
    \item $T_{\rm{60}}$ is defined as the time when the sound energy decays by 60 dB after the source is switched off. It can be computed from the energy decay curve of the RIR.
    \item $C_{\rm{50}}$ measures the energy ratio between early reflections and late reverberation. It can be obtained from RIRs as \cite{C5015}
     \begin{equation}
        C_{50\,m} = 10\log_{10} \frac{\sum_{t=0}^{t_{\rm{d}}+t_{50}}h^2_m(t)}{{{\sum_{t=t_{\rm{d}}+t_{50}}^{\infty}h^2_m(t)}}},
    \label{eq_c50}
    \end{equation}
    where $t_{50}$ is the number of samples for 50 ms. 
    \item The surface area-weighted mean absorption coefficient is computed as \cite{ABS_CNNMLP_fromRIR_JASA21, ABS_T60_SUR_VOL_WASPAA21}
    \begin{equation}
        \bar{\alpha} = \frac{\sum_{i=1}^{I} {S_i\alpha_i}}{\sum_{i=1}^{I}{S_i}},
    \label{eq_abs}
    \end{equation}
    where $I$ is the number of room surfaces, and $S_i$ and $\alpha_i$ represent the surface area and the absorption coefficients of the $i$-th surface, respectively.
\end{itemize}

These spatial acoustic parameters are all continuous values, so we treat spatial acoustic parameter estimation as regression problems. The MSE loss between the predictions and the ground truths is used to train the downstream model.

\subsection{Encoder Model: Multi-Channel Audio Conformer}
\label{sec:mc_conformer}
The model architecture of the spatial encoder is required to be suitable for both the pretext task and downstream tasks.
Existing spatial acoustic parameter estimation works commonly adopt CNN and recurrent neural network (RNN) architectures \cite{TDOA_LSTM_ICASSP19,WYBIS23,DRR_LSTM_IS23,T60_CNN_IWAENC18,ABS_CNNMLP_fromRIR_JASA21,VOL_CNN_ICASSP19,T60_ELR_TASLP19,DRR_T60_SNR_ICASSP20,ABS_T60_SUR_VOL_WASPAA21,T60_SUR_VOL_IWAENC22,VOL_T60_ICASSP23,T60_C50_VOL_ICASSP23,YBTASLP21,T60_CRNN_IS20,SNR_STI_T60_C50_C80_DRR_WASPAA21,T60_C50_JASA23}, in order to leverage the local information modeling ability of convolutional layers and the long-term temporal context learning ability of recurrent layers.
Transformer and Conformer are two widely used architectures for self-supervised learning of audio spectrogram \cite{SSAST22,MAEAST22,ConformerA22,li2023self}.
The Transformer architecture \cite{Transformer17}, known for its ability to capture longer-term temporal context, has outperformed RNN in various audio signal processing tasks \cite{TransRNN19}.
The Conformer \cite{Conformer20} architecture, which incorporates convolutional layers into the Transformer block, has also shown effectiveness in many speech processing tasks \cite{guo2021recent,quan2023spatialnet}.
Therefore, we utilize a combination of CNN and Conformer architectures to construct the encoders in our work.

The spatial acoustic information exhibits discriminative properties in local and global TF regions. Local region means short time and sub band, while global region means long time and full band.
\textbf{Local characteristics}: The CTF model naturally involves local convolution operations along the time axis \cite{talmon2009relative}.
In addition, smoothing over time frames and frequencies is necessary for estimating statistics of random signals, such as the spatial correlation matrix, which will be helpful for the estimation of acoustic parameters \cite{DPD14,FS87,CT08}.
These characteristics motivate us to use time-frequency 2D convolutional layers to capture the local information.
\textbf{Global characteristics}: The spatial acoustic information exhibits certain long-time and full-band properties. This work considers the case that the sound source and microphone array are static, and hence the RIRs remain time-invariant throughout the entire signal.
The directional pulses involving both direct path and early reflections are highly correlated across frequencies. For each propagation path, the frequency-wise IPD is wrapped-linearly increased with respect to frequency \cite{YBTASLP21, YBICASSP22}, and the slope corresponds to the TDOA of this path.
These observations motivate the use of fully connected layers (across all frequencies) to capture the full-band linear dependence, and self-attention scheme to learn the (time-invariant) temporal dependence of spatial acoustic information.

Based on the considerations mentioned above, we design a MC-Conformer as our spatial encoder. Since the CNN and Conformer architectures are widely used for spectral pattern learning, we also use the MC-conformer as our spectral encoder. As shown in Fig.~\ref{fig:meth_network} (a), the MC-Conformer follows a local-to-global processing pipeline. Local processing model utilizes time-frequency 2D convolutional layers to extract short-term and sub-band information. Global processing model uses Conformer blocks to mainly capture long-term and full-band information.
We adopt the sandwich-structured Conformer block presented in \cite{Conformer20}, which stacks a half-step feed-forward module, a multi-head self-attention module, a 1D convolution module, a second half-step feed-forward module and a layernorm layer. The feed-forward module is applied to the time-wise full-band features, which can capture the full-band frequency dependence. The 1D convolution module and multi-head self-attention module mainly learn the short-term and long-term temporal dependencies, respectively.

\section{Experiments and Discussions}
\label{sec:exp}
In this section, we conduct experiments on both simulated data and real-world data to evaluate the effectiveness of the proposed method. We first describe the experimental datasets and configurations, and then present extensive experimental results and discussions.

\subsection{Experimental Setup}
\subsubsection{Simulated dataset}
A large number of rectangular rooms are simulated using an implementation of the image method \cite{Image_method79} provided by gpuRIR toolbox \cite{gpuRIR20}\footnote{\url{https://github.com/DavidDiazGuerra/gpuRIR}}.
The size of simulated rooms ranges from 3$\times$3$\times$2.5 m$^3$ to 15$\times$10$\times$6 m$^3$. The reverberation time is in the range of 0.2 s to 1.3 s. The absorption coefficients of six walls are in [0, 1] and can be totally different. The array has two microphones with an aperture in [3 cm, 20 cm], which is placed parallel to the floor. The center of the microphone array is set in the [20\%, 20\%, 10\%] to [80\%, 80\%, 50\%] of room space in order to ensure a minimum boundary distance from the walls. The static omnidirectional sound source is randomly placed in the room, with a minimum distance of 30 cm from the microphone array and from each room surface.
Speech recordings from the training, development and evaluation subsets of the WSJ0 dataset\footnote{\url{https://catalog.ldc.upenn.edu/LDC93S6A}} are used as source signals, which are used for training, validation and test of the proposed model, respectively.
An arbitrary noise field generator\footnote{\url{https://github.com/ehabets/ANF-Generator}} is used to generate the spatially-diffuse white noise \cite{Diffuse08} for different array apertures.
The microphone signals are generated by first filtering four-second speech recordings with RIRs, and then scaling and adding diffuse noise with a SNR ranging from 15 dB to 30 dB.
Each signal is a random combination of the aforementioned data settings regarding source position, microphone position, source signal, noise signal, room size, reverberation time, relative ratios among six absorption coefficients, etc.

Note that, the proposed pretext task, i.e. cross-channel signal reconstruction, is an ill-posed problem for random noise, as the random samples of noise are unpredictable. Our preliminary experiments demonstrated that more noise lead to more reconstruction error and decreased capability of spatial acoustic learning.
In this experiment, we set the SNR range as 15-30 dB, which does not introduce much adverse effect. Meanwhile, this SNR condition (namely lager than 15 dB) can be satisfied in many real-world scenes for data recording.

We randomly generate 512,000 training signals for the pretext task, in order to imitate that the real-world unlabeled data used for training can be with enough quantity and diversity. The validation set and the test set of the pretext task contain 5,120 signals each.

As for downstream tasks, we generate a series of new room conditions in terms of room size, reverberation time, absorption coefficient and source-microphone position, and 50 RIRs are randomly generated for each room condition. These rooms are divided without overlap to obtain the training, validation and test sets. To imitate a small number of labeled data with limited diversity used for downstream tasks, and to evaluate the influence of data diversity, the number of training (fine-tuning) rooms are set to 2, 4, 8, 16, 32, 64, 128 and 256, respectively. When the number of training rooms is too small, the results will vary a lot from trial to trial (different trials use different training rooms). Therefore, we conduct 16, 8, 4, 2, 1, 1, 1 and 1 trials for these settings of training room number, respectively, and the averaged results over trials are reported. The numbers of validation rooms and test rooms are both 20. Each RIR is convoluted with two, one and four different source signals for training, validation and test, respectively. Accordingly, the numbers of signals for each training, validation and test room are 100, 50 and 200, respectively.

\begin{table}[t]
 \caption{Settings of our selected data from seven public real-world multi-channel datasets}
  \label{tab:real_data}
  \centering
  \renewcommand\arraystretch{1.2}
  \tabcolsep0.014in
  \begin{tabular}{lccccccccccccc}
    \hline
    \hline
    Dataset &\# Room &Microphone Array \\
    \hline
    MIR \cite{MIR14}        &3 &Three 8-channel linear arrays \\
    MeshRIR \cite{Mesh21}   &1 &441 microphones \\
    DCASE \cite{DCASE20}    &9 &A 4-channel tetrahedral array (EM32) \\
    dEchorate \cite{dEchorate21} &11 &Six 5-channel linear arrays \\
    BUTReverb \cite{BUTReverb19} &9 &An 8-channel spherical array\vspace{0.1cm}\\
    \multirow{4}{*}{ACE \cite{ACE16}}
    &\multirow{4}{*}{7} &A 2-channel array (Chromebook),\\
    & &a 3-channel right-angled triangle array (Mobile), \\
    & &an 8-channel linear array (Lin8Ch), \\
    & &a 32-channel spherical array (EM32)\vspace{0.1cm}\\
    \multirow{3}{*}{LOCATA \cite{LOCATA18}}
    &\multirow{3}{*}{1} &A 15-channel planar array (DICIT),\\
    & &a 12-channel robot array (Robot head), \\
    & &a 32-channel spherical array (Eigenmike)\vspace{0.1cm}\\
    MC-WSJ-AV \cite{MC_WSJ_AV05} &3 &Two 8-channel circular arrays \\
    LibriCSS \cite{LibriCSS20} &1 &A 7-channel circular array \\
    {AMIMeeting \cite{AMI05}} &3 &A 8-channel circular array\\
    AISHELL-4 \cite{AISHELL4_21} &10 &A 8-channel circular array \\
    AliMeeting \cite{AliMeeting22} &21 &A 8-channel circular array\\
    RealMAN \cite{RealMAN} &32 &A 32-channel high-precision array \\
    \hline
    \hline
   \end{tabular}
\end{table}

\subsubsection{Real-world datasets}
We collect 11 public real-world multi-channel datasets. Among them, MIR \cite{MIR14}\footnote{\url{https://www.eng.biu.ac.il/gannot/downloads}},
MeshRIR \cite{Mesh21}\footnote{\url{https://zenodo.org/record/5500451}},
DCASE \cite{DCASE20}\footnote{\url{https://zenodo.org/record/6408611}},
dEchorate \cite{dEchorate21}\footnote{\url{https://zenodo.org/record/6576203}},
BUTReverb \cite{BUTReverb19}\footnote{\url{https://speech.fit.vutbr.cz/software/but-speech-fit-reverb-database}} and
ACE \cite{ACE16}\footnote{\url{https://zenodo.org/record/6257551}} provide real-measured multi-channel RIRs. The microphone signals are created by convolving the real-measured RIRs with source signals from WSJ0, and then adding noise with a SNR ranging from 15 dB to 30 dB if noise signals are provided by the corresponding dataset.
LOCATA \cite{LOCATA18}\footnote{\url{https://zenodo.org/record/3630471}}, MC-WSJ-AV \cite{MC_WSJ_AV05}\footnote{\url{https://catalog.ldc.upenn.edu/LDC2014S03}}, LibriCSS \cite{LibriCSS20}\footnote{\url{https://github.com/chenzhuo1011/libri_css}}, AMIMeeting \cite{AMI05}\footnote{\url{https://groups.inf.ed.ac.uk/ami/download/}}, AISHELL-4 \cite{AISHELL4_21}\footnote{\url{https://www.aishelltech.com/aishell_4}}, AliMeeting \cite{AliMeeting22}\footnote{\url{https://www.openslr.org/119}} and RealMAN\cite{RealMAN}\footnote{\url{https://github.com/Audio-WestlakeU/RealMAN}} provide real-recorded multi-channel speech signals.
From the original multi-channel audio recordings, all the two-channel sub-arrays with an aperture in [3 cm, 20 cm] are selected. We only use the data of a single static speaker in our experiments. Table~\ref{tab:real_data} summarizes the settings of selected data from the collected datasets. There are a total of 111 rooms and more than 40 k RIR settings (in terms of room condition, source position, and array position).

For pre-training, we use all the collected real-world datasets. We distinctively generate 512,000, 4,000 and 4,000 signals for training, validation and test, respectively. The importance weight of each dataset for generating data is set according to the number of rooms and the duration of speech recordings in this dataset.
As for downstream tasks, we use the LOCATA dataset for TDOA estimation, and the ACE dataset for the tasks including DRR, $T_{60}$, $C_{50}$ and mean absorption coefficient estimation.
The ACE dataset provides multi-channel measured RIRs and noise signals. The ACE dataset contains 7 different rooms, which however still lacks room diversity, and thence 7-fold cross-validation is adopted. In each fold, we use one room for test, one room for validation and the other five rooms for training. During training (fine-tuning), each training signal is generated on-the-fly as a random combination of RIRs, source signals (from the WSJ0 dataset) and noise signals. A fixed numbers of signals, i.e. 1,000 and 4,000, are generated for validation and test in each fold, respectively.

\subsubsection{Parameter settings}
The sampling rate of signals is 16 kHz. STFT is performed with a window length of 32 ms and a frame shift of 16 ms. The number of frequencies $F$ is 256. For pre-training, the length of microphone signals is set to 4 s, and correspondingly the number of time frames $N$ is 256.
The input STFT coefficients are normalized by dividing the mean value of the magnitude of the first microphone channel. The number of masked time frames is 128, namely half of the number of all frames.
For downstream tasks, the length of microphone signals is set to 1 s for TDOA estimation and 4 s for the other tasks. In addition to the data of single static speaker in the LOCATA dataset, we also use the data of single moving speaker and assume the speaker is static during each one-second segment.

\subsubsection{Model configurations}
\label{sec:model_configuration}
In the encoder, the setting of the convolution block is shown in Fig. \ref{fig:meth_network}. In Conformer blocks, the number of attention heads is 4, the kernel size of convolutional layers is 31, and the expansion factor of feed-forward layers is 4. For the spectral encoder, we use one Conformer block and the embedding dimension $D^{\rm{spec}}$ is set to 512. For the spatial encoder, we use three Conformer blocks and the embedding dimension $D^{\rm{spat}}$ is set to 256. The decoder has two FC layers, where the first layer is with 3072 hidden units and is activated by a rectified linear unit (ReLU), and the second layer outputs the reconstructed signal vector with a dimension of $2FM$.

\subsubsection{Training details}
\label{sec:training_details}
For self-supervised pre-training, the model is trained from scratch using simulated data in the simulated-data experiments,  while in the real-data experiments, the model is initialized with the pre-trained model on simulated data and then trained using the real-world data. We found that the real-world training data (collected from 41 rooms) are not quite sufficient for pre-training, and initializing the model with the pre-trained model of simulated data is helpful for mitigating this problem. We use the Adam optimizer with an initial learning rate 0.001 and a cosine-decay learning rate scheduler. The batch size is set to 128. The maximum number of training epochs is 30. The best model is the one with the minimum validation loss.

For downstream tasks, the pre-trained spatial encoder is fine-tuned using labeled data. The Adam optimizer is used for fine-tuning. The batch size is set to 8 for experiments on simulated data and 16 for experiments on real-world data. Fine-tuning the model with a small amount of labeled data is difficult and unstable in general \cite{BERT19}, so we have carefully designed the fine-tuning scheme. The validation loss is recursively smoothed along the training epochs to reduce its fluctuations.
The initial learning rate is divided by 10 when the smoothed validation loss does not descend with a patience of 10 epochs, and then the training is stopped when the smoothed validation loss does not decrease for another 10 epochs.
For each task, we search the initial learning rate that achieves the smallest smoothed validation loss. The search range of the learning rate is [5e-5, 1e-4, 5e-4, 1e-3] for experiments on simulated data and [1e-4, 1e-3] for experiments on real-world data.
We ensemble the models of the best epoch and its previous four epochs as the final model.

\subsubsection{Evaluation metrics}
All the downstream tasks, i.e. TDOA, DRR, $C_{50}$, $T_{60}$ and mean absorption coefficient estimation, are evaluated with the mean absolute error (MAE) which computes the averaged absolute error between the estimated and ground-truth values over all the test signals.

\subsection{Comparison with Fully Supervised Learning}
As far as we know, this work is the first one to study the self-supervised learning of spatial acoustic information, and there are no self-supervised baseline methods to compare. Therefore, we compare the proposed self-supervised pre-training plus fine-tuning scheme with a fully supervised learning scheme. {In the fully supervised learning scheme, we train the same network architecture as our downstream model (namely the spatial encoder followed by a mean pooling and a linear head) from scratch using labeled data specific to the downstream task.} Training from scratch with a small amount of data is also challenging and unstable, and we employ the same training scheme as described earlier for fine-tuning. This comparison aims to demonstrate the effectiveness of the proposed self-supervised pre-training method.

\subsubsection{Evaluation on simulated data}
\begin{figure*}[t]
  \centering
  \includegraphics[width=1\linewidth]{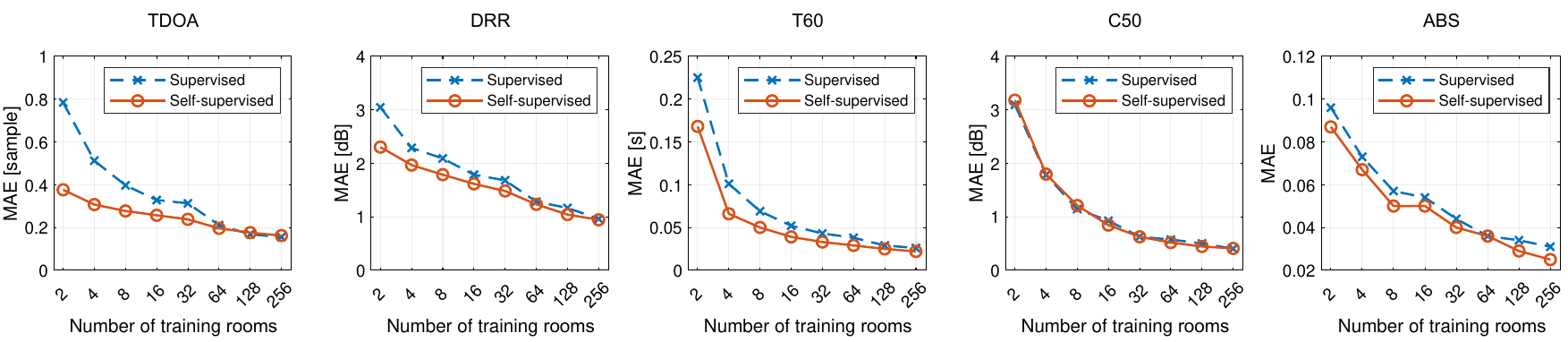}
  \vspace{-0.3cm}
  \caption{Results of TDOA, DRR, $T_{60}$, $C_{50}$ and absorption coefficient (ABS) estimation on the simulated dataset, for the proposed self-supervised pre-training plus fine-tuning method and the fully supervised training method, when using labeled data from different amounts of training rooms. }
  \label{fig:exp_finetune_vs_scratch}
  \vspace{-0.2cm}
\end{figure*}

\begin{figure}[t]
  \centering
  \includegraphics[width=1\linewidth]{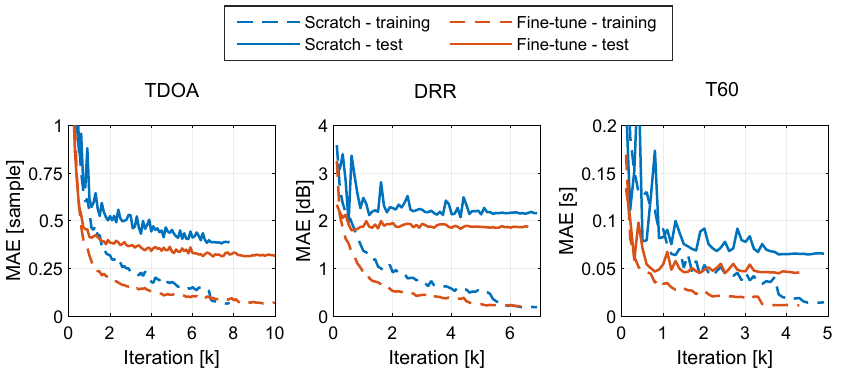}
  \vspace{-0.3cm}
  \caption{Learning curves (MAE versus training iteration) for TDOA, DRR and $T_{60}$ estimation on the simulated dataset, for the proposed fine-tuning scheme and the scheme of training from scratch. The number of training rooms is 8.
  }
  \label{fig:exp_finetune_vs_scratch_ins}
\vspace{-0.1cm}
\end{figure}

Fig.~\ref{fig:exp_finetune_vs_scratch} shows the performance of five spatial acoustic parameter estimation tasks with the two learning schemes when using labeled data from various amounts of training rooms.
It can be observed that the self-supervised setting outperforms the supervised setting under most conditions. This confirms that the spatial encoder learns spatial acoustic information in self-supervised pre-training.
More specifically, the learned representation of relative RIR/CTF involves both the inter-channel information (used for TDOA estimation) and the temporal structure of RIR/CTF (used for DRR, $C_{50}$, $T_{60}$ and mean absorption coefficient estimation).
With the increasing of training rooms (and training data), the MAEs for both self-supervised and supervised settings degrade for all tasks, and the advantage of pre-training becomes less prominent. This highlights the importance of training data diversity, in terms of room conditions, for deep-learning-based acoustic parameter estimation. The significant benefit of pre-training on small labeled datasets confirms that the proposed self-supervised method is promising in real-world applications where data annotation is challenging or resource-consuming. Among the five tasks, one exception is that pre-training is almost not helpful for $C_{\rm{50}}$ estimation, which is possibly because the spatial encoder cannot learn late reverberation well with the current pretext task.

The training and test curves of three downstream tasks in self-supervised and supervised settings are illustrated in Fig.~\ref{fig:exp_finetune_vs_scratch_ins}. Fine-tuning pre-trained models converges faster than training from scratch in general. Although the training losses of the two settings reach a similar level at the end, the test loss of the self-supervised setting is notably lower than the one of the supervised setting. This indicates that pre-training helps to reduce the generalization loss from training to test data.

\begin{table}[t]
 \caption{Performance (MAE) with different training settings on the simulated dataset. The number of training rooms is 8. }
  \label{tab:exp_scratch_linear_finetune_sim}
  \centering
  \renewcommand\arraystretch{1.2}
  \tabcolsep0.07in
  \begin{tabular}{lcccccccccc}
    \hline
    \hline
    \multirow{2}{*}{Training setting} &TDOA &DRR &$T_{60}$ &$C_{50}$ &ABS\\
            &$[$sample$]$ &$[$dB$]$ &$[$s$]$ &$[$dB$]$ &\\
    \hline
    Non-informative                  &2.98 &3.89 &0.287 &4.90 &0.134\\
    Supervised (scratch)       &0.40 &2.09 &0.069 &1.14 &0.057 \\
    \hline
    {Pre-train + linear evaluation}   &1.49 &2.10 &0.083 &1.57 &0.057 \\
    {Pre-train + fine-tune}        &0.28 &1.79 &0.050 &1.21 &0.050 \\
    \hline
    \hline
  \end{tabular}
\end{table}

To evaluate how much information the proposed self-supervised pre-training method has learned, the performance of downstream tasks with four different settings are compared in Table \ref{tab:exp_scratch_linear_finetune_sim}. Non-informative means the acoustic parameter prediction on the test data are simply set as a reasonable non-informative value, namely the mean value of the acoustic parameters of training data, which does not exploit any information from microphone signals of the test dataset. Pre-train plus linear evaluation means the pre-trained model is frozen and only a linear head is trained with downstream data. It can be seen that the linear evaluation setting achieves much better performance measures than the non-informative case, which demonstrates that the pre-trained model/feature indeed involves useful information for downstream tasks. By training/fine-tuning the whole network towards specific downstream tasks, the scratch and fine-tuning settings can better perform on downstream tasks.
Although linear evaluation was once a standard way for evaluating the performance of self-supervised learning methods, it misses the opportunity to pursue strong but non-linear features, which is indeed a strength of deep learning \cite{MAE22}. Therefore, more self-supervised learning works put emphasis on the fine-tuning setting than linear evaluation, and we will also only evaluate the fine-tuning setting in the following.

\begin{table}[t]
 \caption{Performance of the proposed method with different pre-training epochs/iterations}
  \label{tab:exp_pretrain_vs_scratch_epoch}
  \centering
  \renewcommand\arraystretch{1.2}
  \tabcolsep0.12in
  \begin{tabular}{lccccccccccccc}
    \hline
    \hline
    \multirow{2}{*}{\# Epoch / Iteration} &Pre-train. &TDOA  &DRR &$T_{60}$ \\
    &MSE &$[$sample$]$ &$[$dB$]$ &$[$s$]$\vspace{0.03cm}\\
    \hline
    {0 / 0 k (supervised)}    &4.48 &0.40 &2.09 &0.069 \\
    5 / 20 k    &0.84 &0.30 &1.81 &0.059 \\
    10 / 40 k   &0.75 &0.26 &1.81 &0.057 \\
    20 / 80 k   &0.69 &\textbf{0.25} &1.77 &0.053 \\
    \textbf{30 / 120 k}  &\textbf{0.66} &\textbf{0.25} &\textbf{1.74} &\textbf{0.050} \\
    \hline
    \hline
   \end{tabular}
\end{table}

To assess the impact of pre-training epochs/iterations on the performance of downstream tasks, we present the pre-training MSE and the performance of downstream tasks with different pre-training epochs/iterations in Table~\ref{tab:exp_pretrain_vs_scratch_epoch}.
It can be seen that the performance of downstream tasks is consistent with the pretext task to a large extent, namely the performance of downstream tasks can be improved when the pre-training loss is reduced. This property is very important for validating that the proposed pretext task is indeed learning information that can be transferred to downstream tasks.

\subsubsection{Evaluation on real-world data}

\begin{figure}[t]
  \centering
  \includegraphics[width=0.99\linewidth]{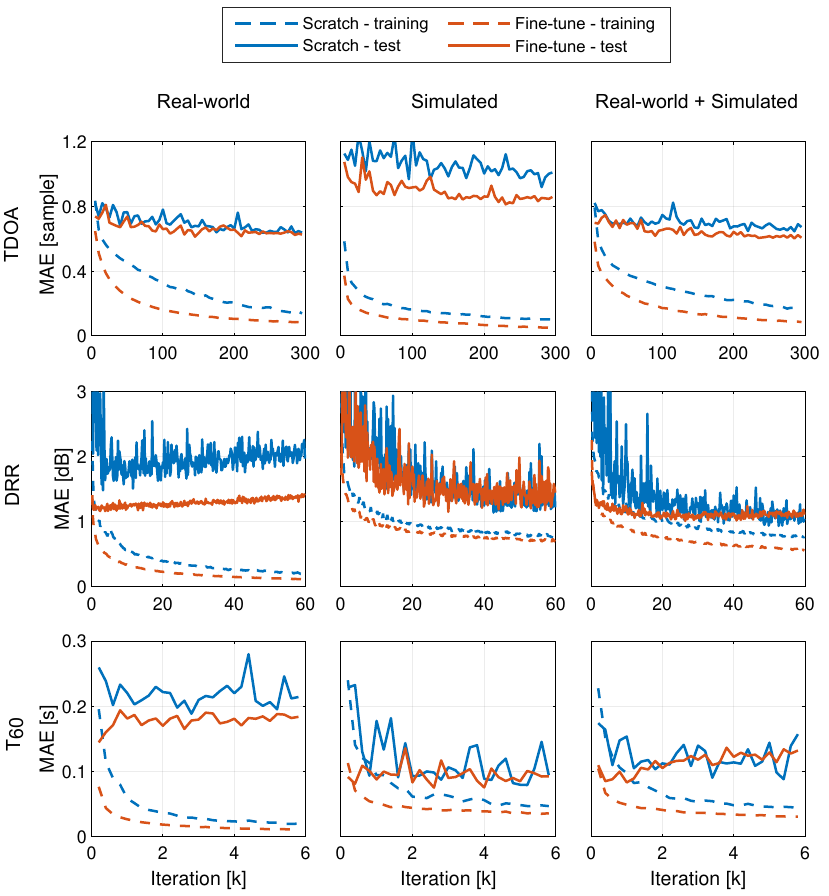}
  \vspace{-0.2cm}
  \caption{Learning curves (MAE versus training iteration) for TDOA, DRR and $T_{60}$ estimation on the real-world datasets, for the proposed fine-tuning scheme and the scheme of training from scratch. The downstream training data is set to real-world data, simulated data, and real-simulated mixed data, respectively. For DRR and $T_{60}$ estimation, the MAEs are the averaged results of 7-fold cross-validation.
  }
  \label{fig:exp_finetune_vs_scratch_realins}
\vspace{-0.1cm}
\end{figure}

We evaluate the proposed self-supervised method on real-world data to validate its effectiveness for practical applications. It is complicated to conduct real-data experiments mainly for two reasons.
One is that we don't have a sufficient amount of real-world data for pre-training, despite the fact that self-supervised pre-training does not require any data annotation. As mentioned in Section \ref{sec:training_details}, we only use collected real-world data of 41 rooms for pre-training, which is not sufficient for fully pre-train the model.
As indicated in Table~\ref{tab:exp_pretrain_vs_scratch_epoch}, the performance of pre-training is closely related to the performance of downstream tasks, so we think the capability of pre-training may not be fully reflected in this experiment.
The other reason is that for fine-tuning or training from scratch in the downstream tasks, it is not necessary to only use a small amount of real data, as a large amount of labeled simulated data can be easily obtained and used. Most DNN-based acoustic parameter estimation methods \cite{TDOA_LSTM_ICASSP19,YBTASLP21,WYBIS23,DRR_LSTM_IS23,T60_CNN_IWAENC18,T60_CRNN_IS20,ABS_CNNMLP_fromRIR_JASA21,VOL_CNN_ICASSP19} train the model (from scratch) using a large amount of labeled simulation data.
Therefore, we conduct experiments of fine-tuning or training from scratch using three groups of data i) a limited number of real-world data; ii) a sufficiently large amount of simulated data generated from 1000 rooms; iii) both real-world and simulated data, and their importance weights are set to 0.5: 0.5. These three settings are evaluated in  Fig.~\ref{fig:exp_finetune_vs_scratch_realins} and Table~\ref{tab:supervision_mode}.

The fine-tuning/training processes of three downstream tasks are illustrated in Fig.~\ref{fig:exp_finetune_vs_scratch_realins}. Note that, different from the training scheme presented in Section \ref{sec:training_details}, to fully plot and analyze the training process in this figure, the learning rate is not reduced and the training is not stopped when it converges.
When using only a small amount of real-world data, fine-tuning the pre-trained model converges rapidly for the DRR and $T_{60}$ estimation tasks, while training from scratch needs more steps to converge. The performance gap between training and test for both self-supervised and supervised settings is very large, indicating that the models largely overfit to the small training dataset. The pre-training helps to mitigate the overfitting, and hence the self-supervised setting achieves better test performance.
When using only simulated data, the performance gaps between training and test vary among the three tasks. TDOA estimation on has the largest gap while $T_{60}$ estimation has the smallest gap. As the quantity and diversity of training data are very large, these performance gaps reflect the simulation-to-reality generalization loss, and different tasks have different losses. Compared with using real-world data, the performance of TDOA estimation gets worse, the performance of $T_{60}$ estimation gets better, and the performance of DRR estimation stays similar.
When using only simulated data, the advantage of the self-supervised setting becomes less prominent, possibly due to the data inconsistency between real-data pre-training and simulated-data fine-tuning.
When using real-simulated mixed data, we hope to combine the advantages of real data and simulated data. Roughly speaking, for each task and for both self-supervised and supervised settings, the real-simulated mixed data can achieve a performance level similar to the better one of pure real data and pure simulated data, but it is hard to surpass the better one.

\begin{table}[t]
  \caption{Performance of spatial acoustic parameter estimation on the real-world datasets}
  \label{tab:supervision_mode}
  \centering
  \renewcommand\arraystretch{1.2}
  \tabcolsep0.04in
  \begin{tabular}{lccccccccccccc}
    \hline
    \hline
    \multirow{2}{*}{Method} &{Downstream} &TDOA &DRR &$T_{60}$ &$C_{50}$ &ABS \\
    &training data &$[$sample$]$ &$[$dB$]$  &$[$s$]$ &$[$dB$]$ &\vspace{0.03cm}\\
    \hline
    \cite{GCCPHAT1976} &- &0.87 & - & - & -  & -\\ 
    \cite{DRR11} &- &- &4.60 & - & -  & -\\ 
    \cite{T6003} &- & - & - &0.179 & -  & -\\ 
    \hline
    \multirow{4}{*}{Supervised}
    &Real-world             &0.65 &1.72 &0.206 &1.20 &0.062\\
    &Simulated              &0.94 &1.03 &0.108 &\textbf{0.78} &0.048\\
    &Simulated (+ real-FT)    &0.66 &\textbf{0.85} &0.152 &0.98 &0.076 \\
    &Real-world + Simulated   &0.66 &1.00 &0.176 &0.85 &0.046\vspace{0.03cm}\\
    \hline
    \multirow{3}{*}{\textbf{\makecell[l]{Self-\\supervised}}}
    &Real-world             &0.65           &1.21 &0.159            &1.30 &0.053\\
    &Simulated              &0.90           &1.35 &\textbf{0.089}   &0.82 &\textbf{0.037}\\
    &Real-world + Simulated   &\textbf{0.64}  &1.13 &0.099            &0.80 &0.042\\
    \hline
    \hline
   \end{tabular}
\vspace{-0.2cm}
\end{table}

Table~\ref{tab:supervision_mode} shows the final performance of the five tasks using the training scheme described in Section \ref{sec:training_details}. As for the supervised case, one extra setting is added, namely supervised training with simulated data and then supervised fine-tuning with real-world data (denoted as Simulated (+ real-FT)). In addition, some conventional methods are also compared, including GCC-PHAT \cite{GCCPHAT1976} for TDOA estimation, one blind DRR estimation method \cite{DRR11} and one blind $T_{60}$ estimation method \cite{T6003}.

The best performance is highlighted in bold for each task. Compared with the supervised setting, the proposed self-supervised setting wins on estimating TDOA, $T_{60}$ and absorption coefficient (ABS), and loses on estimating DRR and $C_{50}$.
Overall, these results on (limited amount of pre-training) real-world data are still promising for showing the effectiveness of self-supervised learning of spatial acoustic information.
We think that it is possible to further improve the capability of self-supervised pre-training when we can record/collect more real data for pre-training, which is not very difficult as it does not require any data annotation.

Compared to conventional methods, the best-performed learning-based models achieve much better performance, which demonstrates the superiority of deep learning for acoustic parameter estimation if the network can be properly trained.

\subsection{Ablation Study}
We conduct some ablation experiments to evaluate the effectiveness of each component of the proposed method. Since existing real-world datasets lack diversity in room conditions, we perform ablation studies on the simulated dataset for better analysis.
The number of simulated training rooms is set to 8 and four trials are performed for each experiment setting unless otherwise stated. Three representative downstream tasks are mainly considered, namely the estimation of TDOA, DRR and $T_{60}$, which depends on the information of direct path, both direct and reflective paths, and reflective paths, respectively.

\subsubsection{Influence of masking rate and comparison with patch-wise scheme}
\begin{table}[t]
 \caption{Performance for the proposed method with different masking rates, and with frame-wise and patch-wise schemes}
  \label{tab:exp_mask}
  \centering
  \renewcommand\arraystretch{1.2}
  \tabcolsep0.1in
  \begin{tabular}{lccccccccccccc}
    \hline
    \hline
    \multirow{2}{*}{Setting} &Masking   &Pre-train.  &TDOA  &DRR  &$T_{60}$\\
    & rate & MSE & $[$sample$]$ & $[$dB$]$ &$[$s$]$\vspace{0.03cm}\\
    \hline
    \multirow{3}{*}{\textbf{Frame-wise}} &25\% &0.34 &0.27 &1.83 &0.052\\
    &{50\%} &0.66 &0.28 &\textbf{1.79} &\textbf{0.050}\\
    &75\% &1.09 &\textbf{0.24} &1.80 &0.055\\
    \hline
    Patch-wise &50\% &1.48 &{0.27} &{1.68} &0.061\\
    \hline
    \hline
   \end{tabular}
\end{table}

Table~\ref{tab:exp_mask} shows the results of three different masking rates, i.e. 25\%, 50\% and 75\%. The pre-training MSE becomes larger with the increase of the masking rate, which is reasonable as reconstructing more frames is more difficult. However, the performance of downstream tasks is comparable for the three masking rates. The masking rate 75\% achieves slightly better TDOA performance, while the masking rate 50\% provides slightly better DRR and $T_{60}$ performance.
Overall, the performance of downstream tasks is not very sensitive to the three values of masking rates, and hence we set the masking rate to the median value 50\% in other experiments.

In many audio spectral pattern learning works \cite{SSAST22,MAEAST22}, the so-called patch-wise scheme outperforms the frame-wise scheme on some downstream tasks, so we also test the patch-wise scheme. Patch-wise means the STFT coefficients are split into patches along the time and frequency axes, and the patches are ranked as a sequence and fed into the Conformer network. In this experiment, the 256 frames $\times$ 256 frequencies are split into 16$\times$16 patches. Note that the frame-wise scheme can be considered as 256$\times$1 patches. The results of the patch-wise scheme with 50\% masking rate are also shown in Table~\ref{tab:exp_mask}. It can be seen that the pretext task with the patch-wise setting is much more challenging, possibly due to that it is difficult to reconstruct 16 continuous frames. For downstream tasks, the patch-wise scheme shows better performance on DRR estimation while worse performance on $T_{60}$ estimation. The reason may be that the frame-wise scheme provides a higher temporal resolution and preserves a finer acoustic reflection structure, which is crucial for $T_{60}$ estimation.

\subsubsection{Contribution of spectral encoder}

\begin{table}[t]
 \caption{Performance of the proposed method with and without using the spectral encoder in pretext and downstream tasks}
  \label{tab:exp_disentangle}
  \centering
  \renewcommand\arraystretch{1.2}
  \tabcolsep0.1in
  \begin{tabular}{llccccccccccc}
    \hline
    \hline
    Pretext &Downstream &TDOA &DRR  &$T_{60}$ \\
    encoder &encoder  &$[$sample$]$ &$[$dB$]$ &$[$s$]$\vspace{0.03cm}\\
    \hline
    W/o  &Spatial         &0.40 &2.09 &0.069\\
    {Spatial}  &Spatial &0.39 &1.80 &0.070\vspace{0.1cm}  \\
    \multirow{3}{*}{\textbf{Spatial+Spectral}} &Spectral         &0.40 &2.02 &0.073 \\
    &\textbf{Spatial}  &0.28 & 1.79 &0.050\\
    &Spatial+Spectral &0.28 &1.75 &0.049\\
    \hline
    \hline
   \end{tabular}
\end{table}

To evaluate the contribution of the spectral encoder, we conduct experiments using the spectral encoder or not in both pretext and downstream tasks. When using only the spatial encoder for the pretext task, the masking scheme given in Eq. (\ref{eq_mask_encoder}) is used, and the encoder is actually required to learn both spatial and spectral information for signal reconstruction. The experimental results are shown in Table~\ref{tab:exp_disentangle}. As a baseline, the performance of training from scratch (namely w/o the pretext encoder) is also given.
Compared with using two encoders for the pretext task and the spatial encoder for downstream tasks, using only one encoder for the pretext task achieves much worse performance on TDOA and $T_{60}$ estimation.
When using two encoders for the pretext task and only the spectral encoder for downstream tasks, similar performance measures are achieved compared with training from scratch. This suggests that the pre-trained spectral encoder does not learn much spatial acoustic information.
Similarly, using both the spectral and spatial encoders for downstream tasks only brings a negligible performance improvement compared to using only the spatial encoder.
Overall, incorporating the spectral encoder in the pretext task to disentangle the learning of signal content makes the spatial encoder more focused on the learning of spatial acoustic information, and hence improves the performance of downstream tasks.

\subsubsection{Comparison of encoder model architectures}
\begin{figure}[t]
  \centering
  \includegraphics[width=1\linewidth]{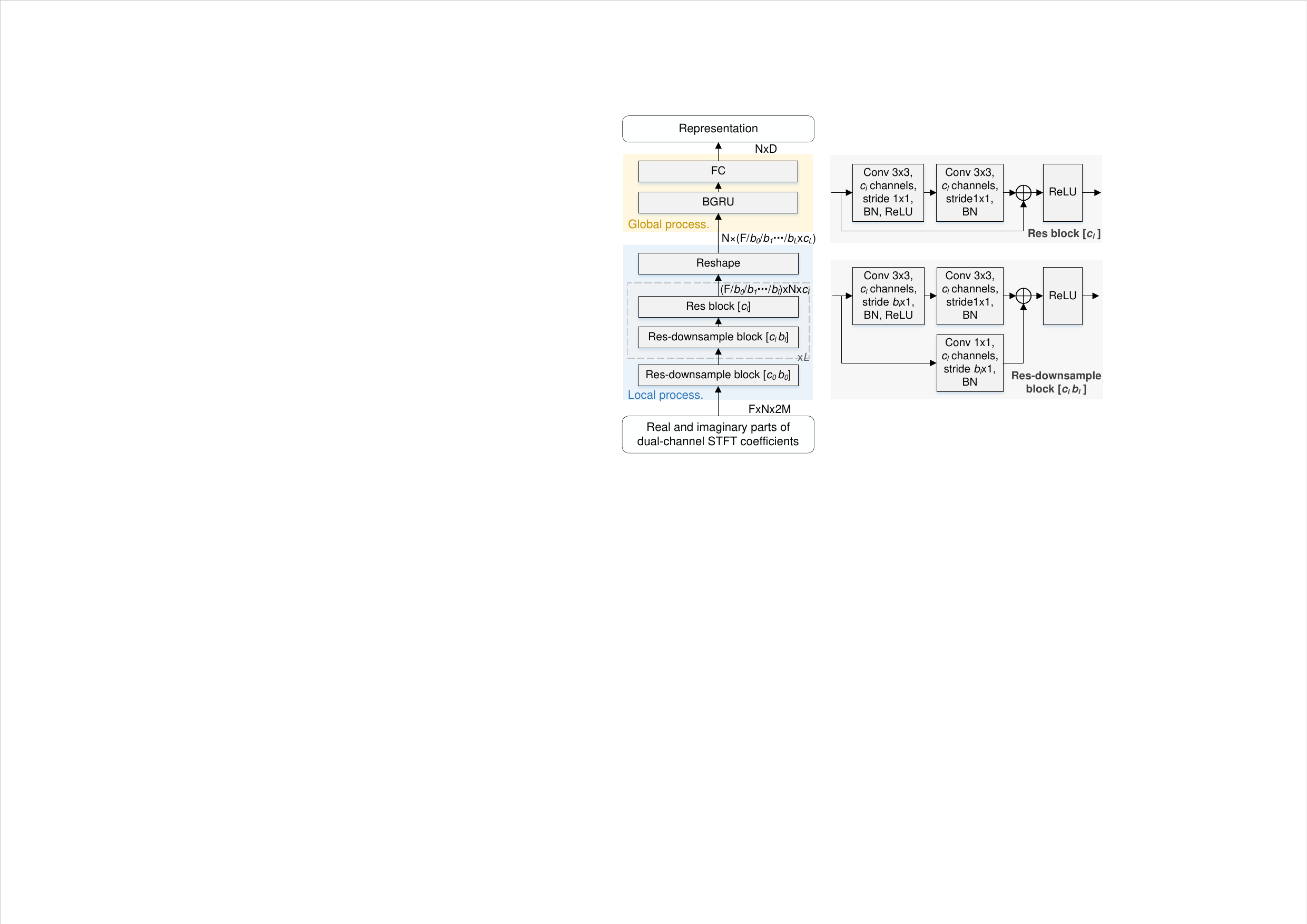}
  \caption{Model architecture of CRNN-based spectral/spatial encoder. }
  \label{fig:meth_network_CRNN}
\end{figure}

\begin{table}[t]
 \caption{Performance of the proposed method with different encoder model architectures}
  \label{tab:exp_network}
  \centering
  \renewcommand\arraystretch{1.2}
  \tabcolsep0.05in
  \begin{tabular}{lccccccccccccc}
    \hline
    \hline
    \multirow{2}{*}{Network} &\# Parameter $[$M$]$ &Pre-train. &TDOA &DRR &$T_{60}$ \\
    &Spectral, Spatial &MSE &$[$sample$]$ &$[$dB$]$ &$[$s$]$\vspace{0.03cm}\\
    \hline
    CRNN &5.4, 2.0 &0.83 &\textbf{0.28} &1.83 &0.064\\
    Transformer \cite{Transformer17} &3.7, 2.6 &1.62 &0.80 &2.36 &0.178\\
    CNN+Transformer &3.8, 2.7 &0.75 &0.30 &1.94 &0.085\\
    Conformer \cite{Conformer20} &6.9, 5.0 &0.82 &0.40 &1.87 &0.056\\
    \textbf{CNN+Conformer} &\multirow{2}{*}{6.9, 5.1} &\multirow{2}{*}{\textbf{0.66}} &\multirow{2}{*}{\textbf{0.28}} &\multirow{2}{*}{\textbf{1.79}} &\multirow{2}{*}{\textbf{0.050}} \\
    \textbf{{(proposed)}} \\
    \hline
    \hline
   \end{tabular}
\end{table}
To demonstrate the effectiveness of the proposed MC-Conformer architecture for both pretext and downstream tasks, we compare the performance of five encoder architectures including CRNN, Transformer, Conformer, CNN+Transformer and CNN+Conformer (namely MC-Conformer).
\begin{itemize}[leftmargin=3.6mm]
    \item CRNN is chosen as CNN and RNN are commonly adopted by spatial acoustic parameter estimation works \cite{TDOA_LSTM_ICASSP19,WYBIS23,DRR_LSTM_IS23,T60_CNN_IWAENC18,ABS_CNNMLP_fromRIR_JASA21,VOL_CNN_ICASSP19,T60_ELR_TASLP19,DRR_T60_SNR_ICASSP20,ABS_T60_SUR_VOL_WASPAA21,T60_SUR_VOL_IWAENC22,VOL_T60_ICASSP23,T60_C50_VOL_ICASSP23,YBTASLP21,T60_CRNN_IS20,SNR_STI_T60_C50_C80_DRR_WASPAA21,T60_C50_JASA23}. The architecture of the CRNN-based encoder is shown in Fig.~\ref{fig:meth_network_CRNN}. It consists of a convolution block to process local TF information and a recurrent block to obtain global TF information. For the spatial encoder, $L=4$, $[c_0,\ldots,c_4]=[16,16,32,64,128]$ and $[b_0,\ldots,b_4]=[1,1,4,4,4]$. For the spectral encoder, $L=2$, $[c_0,c_1,c_2]=[32,32,64]$ and $[b_0,b_1,b_2]=[1,4,4]$. These hyper-parameters have been well tuned to improve the performance of downstream tasks.
    \item Transformer \cite{Transformer17} and Conformer \cite{Conformer20} are widely used in audio/speech processing and self-supervised audio spectrogram learning \cite{SSAST22,MAEAST22,ConformerA22}. We evaluate four types of architectures, i.e., Transformer, Conformer, CNN+Transformer and CNN+Conformer (namely MC-Conformer). The network configurations, including the number of attention heads, the embedding dimension, the number of Transformer/Conformer blocks and the CNN configurations, are set to be the same as the proposed MC-Conformer.
\end{itemize}

The experimental results are shown in Table~\ref{tab:exp_network}. It can be observed that the proposed MC-Conformer outperforms other model architectures in both the pretext and downstream tasks. Transformer alone performs poorly, and once combined with CNN layers (CNN+Transfomer or Conformer) the performance measures improve  significantly. This confirms that CNN is crucial and necessary for capturing local spatial acoustic information. Compared with Conformer, the performance improvement of the proposed CNN+Conformer indicates that using 2D CNNs to pre-process the raw STFT coefficients is very important. Finally, based on these comparisons, we can conclude that CRNN is a strong network architecture for learning spatial acoustic information, while RNN can be replaced with Conformer to better learn long-term dependencies.

\subsection{Qualitative Experiments}
\begin{figure}[t]
  \centering
  \includegraphics[width=0.99\linewidth]{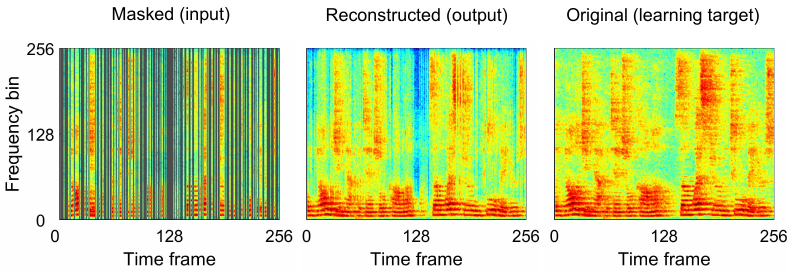}
  \caption{An example of the masked input, the reconstructed signal and the target signal. The reverberation time is 1 s, and the SNR is 20 dB.}
  \label{fig:exp_vis_spectrogram}
\end{figure}

\begin{figure}[t]
  \centering
  \includegraphics[width=0.88\linewidth]{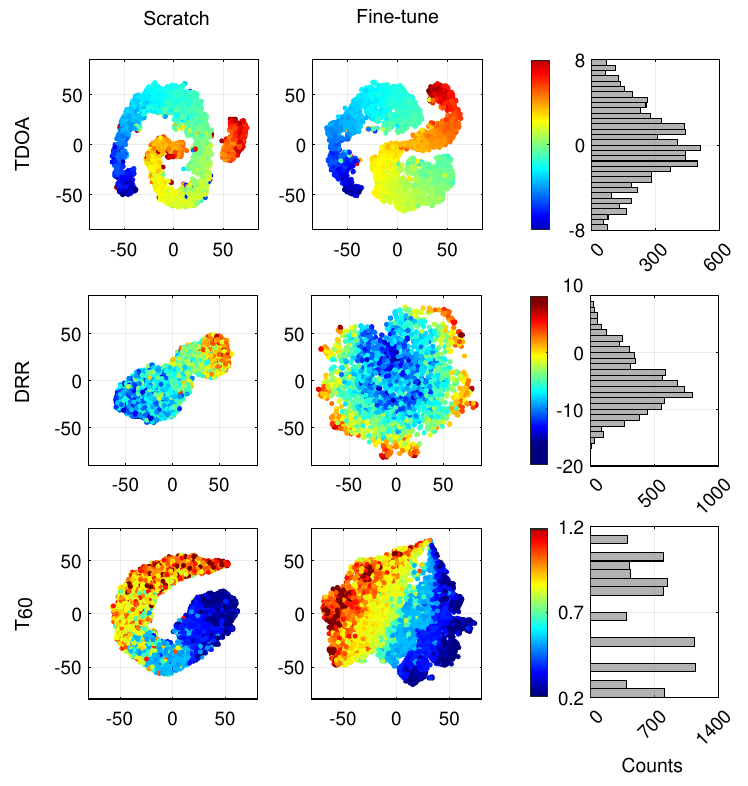}
  \caption{Visualization of the learned representations for three downstream tasks. The number of training rooms is 8. The number of test rooms is 20. The representation extracted after the mean pooling layer from all test data is visualized with the t-SNE technique \cite{tsne}. The gray histograms show the statistics of the values of acoustic parameters in test data. }
  \label{fig:exp_vis_embed}
\end{figure}

Fig.~\ref{fig:exp_vis_spectrogram} provides an example of the reconstructed signal. It can be seen that the main structure of masked frames is well reconstructed.
However, compared with the target signal, the reconstructed signal seems less blurred by reverberation, which is possibly due to that late reverberation has not been well reconstructed. Late reverberation is spatially diffuse with a low spatial correlation, making it more challenging to reconstruct the late reverberation of one channel from that of the other channel. This may be related to the phenomenon that pre-training does not help the estimation of $C_{50}$.

Fig.~\ref{fig:exp_vis_embed} visualizes the learned representations (the hidden vectors after mean pooling) of downstream tasks. Compared with training from scratch, fine-tuning the pre-trained model obtains fewer outliers and presents a much smoother and discriminative manifold. For example, when training from scratch, it is hard to discriminate between the red and yellow points for $T_{60}$ estimation, and between the dark blue and light blue points for DRR estimation. In contrast, they are well discriminated in the fine-tuning results.

\section{Conclusion}
\label{sec:conclusion}
This paper proposes a self-supervised method to learn a universal spatial acoustic representation from dual-channel unlabeled microphone signals. With the designed cross-channel signal reconstruction (CCSR) pretext task, the pretext model is forced to separately learn the spatial acoustic and the spectral pattern information. The dual-encoder plus decoder structure adopted by the pretext task facilitates the disentanglement of the two types of information.
In addition, a novel multi-channel Conformer (MC-Conformer) is utilized to learn the local and global properties of spatial acoustics present in the time-frequency domain, which can boost the performance of both pretext and downstream tasks.
Experiments conducted on both simulated and real-world data verify that the proposed self-supervised pre-training model learns useful knowledge that can be transferred to the spatial acoustics-related tasks including TDOA, DRR, $T_{60}$, $C_{50}$ and mean absorption coefficient estimation. Overall, this work demonstrates the feasibility of learning spatial acoustic information in a self-supervised manner for the first time. Hopefully, this will open a new door for the research of spatial acoustic parameter estimation.

This work mainly focuses on learning spatial acoustic information from dual-channel microphone signals recorded in high-SNR environments with a single static speaker. This acoustic setting can be satisfied in many real-world indoor scenes.
There are several potential directions for future extensions and improvements. For instance, the considered acoustic condition can be more dynamic and complex, and the joint learning of spatial and spectral cues can be further explored.How to extend the proposed method for more than two channels also needs further investigation.

\bibliographystyle{IEEEtran}
\bibliography{mybib}

\begin{thebibliography}{10}
\providecommand{\url}[1]{#1}
\csname url@samestyle\endcsname
\providecommand{\newblock}{\relax}
\providecommand{\bibinfo}[2]{#2}
\providecommand{\BIBentrySTDinterwordspacing}{\spaceskip=0pt\relax}
\providecommand{\BIBentryALTinterwordstretchfactor}{4}
\providecommand{\BIBentryALTinterwordspacing}{\spaceskip=\fontdimen2\font plus
\BIBentryALTinterwordstretchfactor\fontdimen3\font minus
  \fontdimen4\font\relax}
\providecommand{\BIBforeignlanguage}[2]{{%
\expandafter\ifx\csname l@#1\endcsname\relax
\typeout{** WARNING: IEEEtran.bst: No hyphenation pattern has been}%
\typeout{** loaded for the language `#1'. Using the pattern for}%
\typeout{** the default language instead.}%
\else
\language=\csname l@#1\endcsname
\fi
#2}}
\providecommand{\BIBdecl}{\relax}
\BIBdecl

\bibitem{ACE16}
J.~Eaton, N.~D. Gaubitch, A.~H. Moore, and P.~A. Naylor, ``Estimation of room
  acoustic parameters: The {ACE} challenge,'' \emph{IEEE/ACM Trans. Audio,
  Speech, Lang. Process.}, vol.~24, no.~10, pp. 1681--1693, 2016.

\bibitem{GEO_REF_fromRIR21}
W.~Yu and W.~B. Kleijn, ``Room acoustical parameter estimation from room
  impulse responses using deep neural networks,'' \emph{IEEE Trans. Audio,
  Speech, Lang. Process.}, vol.~29, pp. 436--447, 2021.

\bibitem{VR15}
V.~Valimaki, A.~Franck, J.~Ramo, H.~Gamper, and L.~Savioja, ``Assisted
  listening using a headset: Enhancing audio perception in real, augmented, and
  virtual environments,'' \emph{IEEE Signal Process. Mag.}, vol.~32, no.~2, pp.
  92--99, 2015.

\bibitem{Hearingaid01}
J.~M. Kates, ``Room reverberation effects in hearing aid feedback
  cancellation,'' \emph{J. Acoust. Soc. Amer.}, vol. 109, no.~1, pp. 367--378,
  2001.

\bibitem{Robot21}
T.~Zhang, H.~Zhang, X.~Li, J.~Chen, T.~L. Lam, and S.~Vijayakumar,
  ``{AcousticFusion}: Fusing sound source localization to visual {SLAM} in
  dynamic environments,'' in \emph{Proc. IEEE/RSJ Int. Conf. Intell. Robots
  Syst.}, 2021, pp. 6868--6875.

\bibitem{ABS_CNNMLP_fromRIR_JASA21}
C.~Foy, A.~Deleforge, and D.~D. Carlo, ``Mean absorption estimation from room
  impulse responses using virtually supervised learning,'' \emph{J. Acoust.
  Soc. Amer.}, vol. 150, no.~2, pp. 1286--1299, 2021.

\bibitem{RIR_ICASSP22}
A.~Richard, P.~Dodds, and V.~K. Ithapu, ``Deep impulse responses: Estimating
  and parameterizing filters with deep networks,'' in \emph{Proc. IEEE Int.
  Conf. Acoust., Speech, Signal Process.}, 2022, pp. 3209--3213.

\bibitem{RIR_ICASSP23}
A.~Ratnarajah, I.~Ananthabhotla, V.~K. Ithapu, P.~Hoffmann, D.~Manocha, and
  P.~Calamia, ``Towards improved room impulse response estimation for speech
  recognition,'' in \emph{Proc. IEEE Int. Conf. Acoust., Speech, Signal
  Process.}, 2023.

\bibitem{RIREmbed_ICASSP20}
J.~Su, Z.~Jin, and A.~Finkelstein, ``Acoustic matching by embedding impulse
  responses,'' in \emph{Proc. IEEE Int. Conf. Acoust., Speech, Signal
  Process.}, 2020, pp. 426--430.

\bibitem{T60_C50_VOL_ICASSP23}
P.~Gotz, C.~Tuna, A.~Walther, and E.~A.~P. Habets, ``Contrastive representation
  learning for acoustic parameter estimation,'' in \emph{Proc. IEEE Int. Conf.
  Acoust., Speech, Signal Process.}, 2023.

\bibitem{T60_SUR_VOL_IWAENC22}
P.~Srivastava, A.~Deleforge, and E.~Vincent, ``Realistic sources, receivers and
  walls improve the generalisability of virtually-supervised blind acoustic
  parameter estimators,'' in \emph{Proc. Int. Workshop Acoust. Signal
  Enhancement}, 2022.

\bibitem{Diffuse_ISM10}
E.~A. Lehmann and A.~M. Johansson, ``Diffuse reverberation model for efficient
  image-source simulation of room impulse responses,'' \emph{IEEE Trans. Audio,
  Speech, Lang. Process.}, vol.~18, no.~6, pp. 1429--1439, 2010.

\bibitem{gpuRIR20}
D.~Diaz-Guerra, A.~Miguel, and J.~R. Beltran, ``{gpuRIR}: A python library for
  room impulse response simulation with {GPU} acceleration,'' \emph{Multimedia
  Tools Appl.}, vol.~80, no.~4, pp. 5653--5671, 2021.

\bibitem{Conformer20}
A.~Gulati, J.~Qin, C.-C. Chiu, N.~Parmar, Y.~Zhang, J.~Yu, W.~Han, S.~Wang,
  Z.~Zhang, Y.~Wu, and R.~Pang, ``Conformer: Convolution-augmented transformer
  for speech recognition,'' in \emph{Proc. INTERSPEECH}, 2020, pp. 5036--5040.

\bibitem{CNN-Conformer21}
Y.~Zhang, S.~Wang, Z.~Li, K.~Guo, S.~Chen, and Y.~Pang, ``Data augmentation and
  class-based ensembled {CNN-Conformer} networks for sound event localization
  and detection,'' in \emph{Detect. and Classification of Acoust. Scenes Events
  Workshop Technical Report}, 2021.

\bibitem{TDOA_LSTM_ICASSP19}
P.~Pertila and M.~Parviainen, ``Time difference of arrival estimation of speech
  signals using deep neural networks with integrated time-frequency masking,''
  in \emph{Proc. IEEE Int. Conf. Acoust., Speech, Signal Process.}, 2019, pp.
  436--440.

\bibitem{YBTASLP21}
B.~Yang, H.~Liu, and X.~Li, ``Learning deep direct-path relative transfer
  function for binaural sound source localization,'' \emph{IEEE/ACM Trans.
  Audio, Speech, Lang. Process.}, vol.~29, pp. 3491--3503, 2021.

\bibitem{WYBIS23}
Y.~Wang, B.~Yang, and X.~Li, ``{FN-SSL}: Full-band and narrow-band fusion for
  sound source localization,'' in \emph{Proc. INTERSPEECH}, 2023, pp.
  3779--3783.

\bibitem{DRR_LSTM_IS23}
W.~Mack, S.~Deng, and E.~A.~P. Habets, ``Single-channel blind
  direct-to-reverberation ratio estimation using masking,'' in \emph{Proc.
  INTERSPEECH}, 2020, pp. 5066--5070.

\bibitem{T60_CNN_IWAENC18}
H.~Gamper and I.~J. Tashev, ``Blind reverberation time estimation using a
  convolutional neural network,'' in \emph{Proc. Int. Workshop Acoust. Signal
  Enhancement}, 2018, pp. 136--140.

\bibitem{T60_CRNN_IS20}
S.~Deng, W.~Mack, and E.~A.~P. Habets, ``Online blind reverberation time
  estimation using {CRNNs},'' in \emph{Proc. INTERSPEECH}, 2020, pp.
  5061--5065.

\bibitem{VOL_CNN_ICASSP19}
A.~F. Genovese, H.~Gamper, V.~Pulkki, N.~Raghuvanshi, and I.~J. Tashev, ``Blind
  room volume estimation from single-channel noisy speech,'' in \emph{Proc.
  IEEE Int. Conf. Acoust., Speech, Signal Process.}, 2019, pp. 231--235.

\bibitem{T60_ELR_TASLP19}
F.~Xiong, S.~Goetze, B.~Kollmeier, and B.~T. Meyer, ``Joint estimation of
  reverberation time and early-to-late reverberation ratio from single-channel
  speech signals,'' \emph{IEEE/ACM Trans. Audio, Speech, Lang. Process.},
  vol.~27, no.~2, pp. 255--267, 2019.

\bibitem{DRR_T60_SNR_ICASSP20}
D.~Looney and N.~D. Gaubitch, ``Joint estimation of acoustic parameters from
  single-microphone speech observations,'' in \emph{Proc. IEEE Int. Conf.
  Acoust., Speech, Signal Process.}, 2020, pp. 431--435.

\bibitem{SNR_STI_T60_C50_C80_DRR_WASPAA21}
P.~C. Paula Sanchez L~opez and M.~Cernak, ``A universal deep room acoustics
  estimator,'' in \emph{Proc. IEEE Workshop Appl. Signal Process. Audio
  Acoust.}, 2021, pp. 356--360.

\bibitem{ABS_T60_SUR_VOL_WASPAA21}
P.~Srivastava, A.~Deleforge, and E.~Vincent, ``Blind room parameter estimation
  using multiple multichannel speech recordings,'' in \emph{Proc. IEEE Workshop
  Appl. Signal Process. Audio Acoust.}, 2021, pp. 226--230.

\bibitem{VOL_T60_ICASSP23}
C.~Ick, A.~Mehrabi, and W.~Jin, ``Blind acoustic room parameter estimation
  using phase features,'' in \emph{Proc. IEEE Int. Conf. Acoust., Speech,
  Signal Process.}, 2023.

\bibitem{T60_C50_JASA23}
P.~Gotz, C.~Tuna, A.~Walther, and E.~A.~P. Habet, ``Online reverberation time
  and clarity estimation in dynamic acoustic conditions,'' \emph{J. Acoust.
  Soc. Amer.}, vol. 153, no.~6, pp. 3532--3542, 2023.

\bibitem{SSL22}
L.~Ericsson, H.~Gouk, C.~C. Loy, and T.~M. Hospedales, ``Self-supervised
  representation learning: Introduction, advances, and challenges,'' \emph{IEEE
  Signal Process. Mag.}, vol.~39, no.~3, pp. 42--62, 2022.

\bibitem{SSL_speech22}
A.~Mohamed, H.~yi~Lee, L.~Borgholt, J.~D. Havtorn, J.~Edin, C.~Igel,
  K.~Kirchhoff, S.-W. Li, K.~Livescu, L.~Maaløe, T.~N. Sainath, and
  S.~Watanabe, ``Self-supervised speech representation learning: A review,''
  \emph{IEEE J. Selected Topics Signal Process.}, vol.~16, no.~6, pp.
  1179--1210, 2022.

\bibitem{CPC18}
A.~van~den Oord, Y.~Li, and O.~Vinyals, ``Representation learning with
  contrastive predictive coding,'' \emph{arXiv preprint arXiv:1807.03748},
  2018.

\bibitem{wav2vec19}
S.~Schneider, A.~Baevski, R.~Collobert, and M.~Auli, ``wav2vec: Unsupervised
  pre-training for speech recognition,'' in \emph{Proc. INTERSPEECH}, 2019, pp.
  3465--3469.

\bibitem{COLA21}
A.~Saeed1, D.~Grangier, and N.~Zeghidour, ``Contrastive learning of
  general-purpose audio representations,'' in \emph{Proc. IEEE Int. Conf.
  Acoust., Speech, Signal Process.}, 2021, pp. 3875--3879.

\bibitem{BYOLA23}
D.~Niizumi, D.~Takeuchi, Y.~Ohishi, N.~Harada, and K.~Kashino, ``{BYOL} for
  audio: Exploring pre-trained general-purpose audio representations,''
  \emph{IEEE/ACM Trans. Audio, Speech, Lang. Process.}, vol.~31, pp. 137--151,
  2023.

\bibitem{APC19}
Y.-A. Chung, W.-N. Hsu, H.~Tang, and J.~Glass, ``An unsupervised autoregressive
  model for speech representation learning,'' in \emph{Proc. INTERSPEECH},
  2019, pp. 146--150.

\bibitem{APC20}
Y.-A. Chung and J.~Glass, ``Generative pre-training for speech with
  autoregressive predictive coding,'' in \emph{Proc. IEEE Int. Conf. Acoust.,
  Speech, Signal Process.}, 2020, pp. 3497--3501.

\bibitem{BERT19}
J.~Devlin, M.-W. Chang, K.~Lee, and K.~Toutanova, ``{BERT}: Pre-training of
  deep bidirectional transformers for language understanding,'' in \emph{Proc.
  Conf. North Amer. Chapter Assoc. Comput. Linguistics: Hum. Lang. Technol.},
  2019, pp. 4171--4186.

\bibitem{SSAST22}
Y.~Gong, C.-I.~J. Lai, Y.-A. Chung, and J.~Glass, ``{SSAST}: Self-supervised
  audio spectrogram transformer,'' in \emph{Proc. AAAI Conf. Artif. Intell.},
  2022, pp. 10\,699--10\,709.

\bibitem{MAEAST22}
A.~Baade, P.~Peng, and D.~Harwath, ``{MAE-AST}: Masked autoencoding audio
  spectrogram transformer,'' in \emph{Proc. INTERSPEECH}, 2022, pp. 2438--2442.

\bibitem{SepLocSCM23}
H.~Munakata, Y.~Bando, R.~Takeda, K.~Komatani, and M.~Onishi, ``Joint
  separation and localization of moving sound sources based on neural full-rank
  spatial covariance analysis,'' \emph{IEEE Signal Process. Lett.}, vol.~30,
  pp. 384--388, 2023.

\bibitem{SepUnsup23}
Z.-Q. Wang and S.~Watanabe, ``Unssor: Unsupervised neural speech separation by
  leveraging over-determined training mixtures,'' in \emph{Proc. Int. Conf.
  Neural Inf. Process. Syst.}, 2023, pp. 1--22.

\bibitem{talmon2009relative}
R.~Talmon, I.~Cohen, and S.~Gannot, ``Relative transfer function identification
  using convolutive transfer function approximation,'' \emph{IEEE Trans. Audio,
  Speech, Lang. Process.}, vol.~17, no.~4, pp. 546--555, 2009.

\bibitem{C5015}
P.~P. Parada1, D.~Sharma, P.~A. Naylor, and T.~van Waterschoot, ``Reverberant
  speech recognition exploiting clarity index estimation,'' \emph{EURASIP J.
  Advances Signal Process.}, vol. 2015, no.~54, pp. 1--12, 2015.

\bibitem{ConformerA22}
S.~Srivastava, Y.~Wang, A.~Tjandra, A.~Kumar, C.~Liu, K.~Singh, and Y.~Saraf,
  ``Conformer-based self-supervised learning for non-speech audio tasks,'' in
  \emph{Proc. IEEE Int. Conf. Acoust., Speech, Signal Process.}, 2022, pp.
  8862--8866.

\bibitem{li2023self}
X.~Li, N.~Shao, and X.~Li, ``Self-supervised audio teacher-student transformer
  for both clip-level and frame-level tasks,'' \emph{IEEE/ACM Trans. Audio,
  Speech, Lang. Process.}, vol.~32, pp. 1336--1351, 2024.

\bibitem{Transformer17}
A.~Vaswani, N.~Shazeer, N.~Parmar, J.~Uszkoreit, L.~Jones, A.~N. Gomez, Łukasz
  Kaiser, and I.~Polosukhin, ``Attention is all you need,'' in \emph{Proc. Int.
  Conf. Neural Inf. Process. Syst.}, 2017, pp. 5998--6008.

\bibitem{TransRNN19}
S.~Karita1, N.~Chen, T.~Hayashi, T.~Hori, and et~al, ``A comparative study on
  {Transformer} vs {RNN} in speech applications,'' in \emph{Proc. IEEE Autom.
  Speech Recognit. Understanding Workshop}, 2019, pp. 449--456.

\bibitem{guo2021recent}
P.~Guo, F.~Boyer, X.~Chang, T.~Hayashi, Y.~Higuchi, H.~Inaguma, N.~Kamo, C.~Li,
  D.~Garcia-Romero, J.~Shi \emph{et~al.}, ``Recent developments on {ESPnet}
  toolkit boosted by {Conformer},'' in \emph{Proc. IEEE Int. Conf. Acoust.,
  Speech, Signal Process.}, 2021, pp. 5874--5878.

\bibitem{quan2023spatialnet}
C.~Quan and X.~Li, ``{SpatialNet}: Extensively learning spatial information for
  multichannel joint speech separation, denoising and dereverberation,''
  \emph{IEEE/ACM Trans. Audio, Speech, Lang. Process.}, vol.~32, pp.
  1310--1323, 2024.

\bibitem{DPD14}
O.~Nadiri and B.~Rafaely, ``Localization of multiple speakers under high
  reverberation using a spherical microphone array and the direct-path
  dominance test,'' \emph{IEEE/ACM Trans. Audio, Speech, Lang. Process.},
  vol.~22, no.~10, pp. 1494--1505, 2014.

\bibitem{FS87}
H.~Wang, C.~C. Li, and J.~X. Zhu, ``High-resolution direction finding in the
  presence of multipath: A frequency-domain smoothing approach,'' in
  \emph{Proc. IEEE Int. Conf. Acoust., Speech, Signal Process.}, 1987, pp.
  2276--2279.

\bibitem{CT08}
S.~Mohan, M.~E. Lockwood, M.~L. Kramer, and D.~L. Jones, ``Localization of
  multiple acoustic sources with small arrays using a coherence test,''
  \emph{J. Acoust. Soc. Amer.}, vol. 123, no.~4, pp. 2136--2147, 2008.

\bibitem{YBICASSP22}
B.~Yang, H.~Liu, and X.~Li, ``{SRP-DNN}: Learning direct-path phase difference
  for multiple moving sound source localization,'' in \emph{Proc. IEEE Int.
  Conf. Acoust., Speech, Signal Process.}, 2022, pp. 721--725.

\bibitem{Image_method79}
J.~B. Allen and D.~A. Berkley, ``Image method for efficiently simulating
  small-room acoustics,'' \emph{J. Acoust. Soc. Amer.}, vol.~65, no.~4, pp.
  943--950, 1979.

\bibitem{Diffuse08}
E.~A.~P. Habets, I.~Cohen, and S.~Gannot, ``Generating nonstationary
  multisensor signals under a spatial coherence constraint,'' \emph{J. Acoust.
  Soc. Amer.}, vol. 124, no.~5, pp. 2911--2917, 2008.

\bibitem{MIR14}
E.~Hadad, F.~Heese, P.~Vary, and S.~Gannot, ``Multichannel audio database in
  various acoustic environments,'' in \emph{Proc. Int. Workshop Acoust. Signal
  Enhancement}, 2014, pp. 313--317.

\bibitem{Mesh21}
S.~Koyama, T.~Nishida, K.~Kimura, T.~Abe, N.~Veno, and J.~Brunnstrom,
  ``{MeshRIR}: A dataset of room impulse responses on meshed grid points for
  evaluating sound field analysis and synthesis methods,'' in \emph{Proc. IEEE
  Workshop Appl. Signal Process. Audio Acoust.}, 2021, pp. 151--155.

\bibitem{DCASE20}
A.~Politis, S.~Adavanne, and T.~Virtanen, ``A dataset of reverberant spatial
  sound scenes with moving sources for sound event localization and
  detection,'' in \emph{Proc. Detect. and Classification of Acoust. Scenes
  Events Workshop}, 2020, pp. 165--169.

\bibitem{dEchorate21}
D.~D. Carlo1, P.~Tandeitnik, CeFoy, N.~Bertin, A.~Deleforge, and S.~Gannot,
  ``{dEchorate}: A calibrated room impulse response dataset for echo-aware
  signal processing,'' \emph{EURASIP J. Audio, Speech, Music Process.}, vol.
  2021, no.~39, pp. 1--15, 2021.

\bibitem{BUTReverb19}
I.~Szoke, M.~Skacel, L.~Mosner, J.~Paliesek, and J.~H. Cernocky, ``Building and
  evaluation of a real room impulse response dataset,'' \emph{IEEE J. Selected
  Topics Signal Process.}, vol.~13, no.~4, pp. 863--876, 2019.

\bibitem{LOCATA18}
H.~W. Lollmann, C.~Evers, A.~Schmidt, H.~Mellmann, H.~Barfuss, P.~A. Naylor,
  and W.~Kellermann, ``The {LOCATA }challenge data corpus for acoustic source
  localization and tracking,'' in \emph{Proc. IEEE Sensor Array Multichannel
  Signal Process. Workshop}, 2018, pp. 410--414.

\bibitem{MC_WSJ_AV05}
M.~Lincoln, I.~McCowan, J.~Vepa, and H.~K. Maganti, ``The multi-channel wall
  {Street Journal} audio visual corpus {(MC-WSJ-AV)}: specification and initial
  experiments,'' in \emph{Proc. IEEE Autom. Speech Recognit. Understanding
  Workshop}, 2005, pp. 357--362.

\bibitem{LibriCSS20}
Z.~Chen, T.~Yoshioka, L.~Lu, T.~Zhou, Z.~Meng, Y.~Luo, J.~Wu, X.~Xiao, and
  J.~Li, ``Continuous speech separation: Dataset and analysis,'' in \emph{Proc.
  IEEE Int. Conf. Acoust., Speech, Signal Process.}, 2020, pp. 7284--7288.

\bibitem{AMI05}
I.~McCowan, J.~Carletta, W.~Kraaij, S.~Ashby, S.~Bourban, M.~Flynn,
  M.~Guillemot, T.~Hain, J.~Kadlec, V.~Karaiskos, M.~Kronenthal, G.~Lathoud,
  M.~Lincoln, A.~Lisowska, W.~Post, D.~Reidsma, and P.~Wellner, ``The {AMI}
  meeting corpus,'' in \emph{Proc. Int. Conf. Methods Tech. Behav. Res.}, 2005.

\bibitem{AISHELL4_21}
Y.~Fu, L.~Cheng, S.~Lv, Y.~Jv, Y.~Kong, Z.~Chen, Y.~Hu, L.~Xie, J.~Wu, H.~Bu,
  X.~Xu, J.~Du, and J.~Chen, ``Aishell-4: An open source dataset for speech
  enhancement, separation, recognition and speaker diarization in conference
  scenario,'' in \emph{Proc. INTERSPEECH}, 2021, pp. 3665--3669.

\bibitem{AliMeeting22}
F.~Yu, S.~Zhang, Y.~Fu, L.~Xie, S.~Zheng, Z.~Du, W.~Huang, P.~Guo, Z.~Yan,
  B.~Ma, X.~Xu, and H.~Bu, ``M2met: The icassp 2022 multi-channel multi-party
  meeting transcription challenge,'' in \emph{Proc. IEEE Int. Conf. Acoust.,
  Speech, Signal Process.}, 2022, pp. 6167--6171.

\bibitem{RealMAN}
B.~Yang, C.~Quan, Y.~Wang, P.~Wang, Y.~Yang, Y.~Fang, N.~Shao, H.~Bu, X.~Xu,
  and X.~Li, ``{RealMAN}: A real-recorded and annotated microphone array
  dataset for dynamic speech enhancement and localization,'' in \emph{arXiv
  preprint arXiv:2406.19959}, 2024.

\bibitem{MAE22}
K.~He, X.~Chen, S.~Xie, Y.~Li, P.~Dollar, and R.~Girshick, ``Masked
  autoencoders are scalable vision learners,'' in \emph{IEEE/CVF conference on
  computer vision and pattern recognition (CVPR)}, 2022, pp. 15\,979--15\,988.

\bibitem{GCCPHAT1976}
C.~H. Knapp and G.~C. Carter, ``The generalized correlation method for
  estimation of time delay,'' \emph{IEEE Trans. Acoust., Speech, Signal
  Process.}, vol.~24, no.~4, pp. 320--327, 1976.

\bibitem{DRR11}
M.~Jeub, C.~N. v~Christophe~Beaugeant, and P.~Vary, ``Blind estimation of the
  coherent-to-diffuse energy ratio from noisy speech signals,'' in \emph{Proc.
  Euro. Signal Process. Conf.}, 2011, pp. 1347--1351.

\bibitem{T6003}
R.~Ratnam, D.~L. Jones, B.~C. Wheeler, J.~William D.~O’Brien, C.~R. Lansing,
  and A.~S. Feng, ``Blind reverberation time estimation,'' \emph{J. Acoust.
  Soc. Amer.}, vol. 114, no.~5, p. 2877–2892, 2003.

\bibitem{tsne}
L.~V.~D. Maaten and G.~Hinton, ``Visualizing data using {t-SNE},'' \emph{J.
  Mach. Learn. Res.}, vol.~9, pp. 2579--2605, 2008.

\end{thebibliography}

\newpage

%



\vfill

\end{document}